\documentclass[useAMS,usenatbib]{mn2e}

\usepackage{units}
\usepackage{amsmath}
\usepackage{units}
\usepackage{threeparttable}
\usepackage{subfigure}
\usepackage{graphicx}
\usepackage{amssymb}
\usepackage{booktabs}
\usepackage{color} 
\usepackage{ulem}

\providecommand{\apj}{ApJ}
\providecommand{\apjl}{ApJL}

\providecommand{\aap}{A{\&}A}

\providecommand{\araa}{ARA{\&}A}

\providecommand{\mnras}{MNRAS}

\def\MBE{M_{\rm BE}}
\def\Mc{M_{\rm c}}

\newcommand{\beq}{\begin{equation}}
\newcommand{\cs}{c_{\rm s}}
\newcommand{\eeq}{\end{equation}}
\newcommand{\hcop}{HCO$^+$}
\newcommand{\kms}{{\rm km\ s}^{-1}}
\newcommand{\Lb}{L_{\rm b}}
\newcommand{\LJ}{L_{\rm J}}
\newcommand{\MJ}{M_{\rm J}}
\newcommand{\Msun}{M_{\odot}}
\newcommand{\nhp}{N$_2$H$^+$}
\newcommand{\nht}{NH$_3$}
\newcommand{\pcc}{{\rm cm}^{-3}}

\newcommand{\Rminifty}{R_{-\infty}}
\newcommand{\Rshu}{R_{\rm Shu}}
\newcommand{\tff}{t_{\rm ff}}

\newcommand{\vmaxt}{v_{\rm max}(t)}

\title[Synthetic Spectral Signatures from an Isothermal Collapsing Core]{Synthetic Spectral Signatures from Isothermal Collapsing Gas and the Interpretation 
of Infall Profiles}
\author[R. M.~Loughnane et al.]{R. M. Loughnane$^{1}$\thanks{E-mail: rloughnane@irya.unam.mx (RML)}, Enrique V\'{a}zquez-Semadeni$^{1}$\thanks{E-mail: e.vazquez@irya.unam.mx (EVS)}, 
and Ra\'ul Naranjo-Romero$^{1}$\\
$^{1}$Instituto de Radioastronom\'{a} y Astrof\'{i}sica, Universidad Nacional Aut\'{o}noma de M\'{e}xico,
Apdo. Postal 3-72, Morelia, \\
Michoac\'{a}n, 58089, M\'{e}xico\\}

\makeatletter
\g@addto@macro\bfseries{\boldmath}
\makeatother

\begin{document}

\date{\today}

\pagerange{\pageref{firstpage}--\pageref{lastpage}} \pubyear{}

\maketitle

\raggedbottom

\label{firstpage}

\begin{abstract}

We revisit the interpretation of blue-excess molecular-line profiles
from dense collapsing cores, considering recent numerical results
suggesting that the prestellar stage of core collapse occurs from the
outside-in, rather than inside-out. We thus perform synthetic
molecular-line observations of simulated collapsing,
spherically-symmetric, density fluctuations of low initial amplitude,
embedded in a uniform, globally gravitationally unstable background,
without a turbulent component. The collapsing core develops a flattened, 
Bonnor-Ebert-like density profile, but with an outside-in radial velocity 
profile, where the maximum infall speeds occur at large radii, where the 
density is already decreasing, with cloud-to-core accretion, and no 
hydrostatic outer envelope. Using the optically thick \hcop\ $J=1-0$ and 
$3-2$ rotational transitions, we consider several ``typical''
beamwidths and use a simple line-fitting model to infer infall speeds from the
synthetic profiles similarly to how it is done by standard line modeling. 
We find that the model-derived infall speeds are up to $3-4$ times smaller
than the actual peak infall speeds, because the largest infall speeds
are downweighted by the low density of the gas in which they occur,
due to the outside-in nature of the actual radial infall speed profile.
Also, using the isolated N$_{\rm2}$H$^+$ $J_{\rm F_1F}=1_{01}-0_{12}$ 
hyperfine component, we investigate the variation in the so-called asymmetry 
parameter, $\delta v\equiv (V_{\rm thick}-V_{\rm thin})/\Delta v_{\rm thin}$, 
during the collapse, finding good agreement with observed values. Finally, the
\hcop\ $J=3-2$ spectra exhibit extreme $T_{\rm b}$/$T_{\rm r}$--ratios similar 
to those observed in evolved cores, for the larger beamwidths late in the 
collapse. Our results suggest that standard dynamical infall reproduces several
features from observations, but that low-mass core infall speeds are
generally underestimated, often interpreted as being subsonic even when the 
actual speeds are supersonic, due to the incorrect assumption of an inside-out
infall radial velocity profile with a static outer envelope.

\end{abstract}


\section{Introduction} 
\label{sec:intro}


\subsection{The molecular-cloud background of collapsing cores}
\label{sec:backgd}

The process by which dense molecular cloud (MC) cores collapse to form
stars is still an unsettled issue. Under the old picture of magnetically-supported 
clouds with collapse mediated by ambipolar diffusion 
\citep[e.g.,] [] {shu87, Mousch91}, MCs were assumed to be Jeans unstable, 
but generally had magnetically subcritical mass-to-magnetic flux ratios, 
so that they were globally supported by the magnetic field against their 
self-gravity. Dense cores would grow quasistatically as ambipolar diffusion 
allowed local condensations of neutral material to gradually permeate 
through the magnetic field lines, until the local mass-to-flux ratio became 
supercritical, and dynamical collapse could proceed.

However, several issues with that picture emerged at the turn of the
century \citep[see] [for a review] {MK04}, not the least of which is the
realization that MCs are, in general, magnetically supercritical
\citep{Crutcher+10}, and therefore not globally supported by the
magnetic field. Instead, the current prevalent ``gravoturbulent'' view is 
that MCs are supersonically turbulent, and globally supported against 
collapse by virialized turbulent motions \citep[e.g.,] [] {VS+03, MK04}, 
while simultaneously this turbulence induces small-scale density 
fluctuations in which the local Jeans mass may become smaller
than the fluctuation's own mass, and therefore the latter begins to
undergo gravitational collapse \citep[see, e.g., the reviews by] []
{MK04, BP+07, MO07}.

Somewhat surprisingly, notions about dense MC cores that
had originated in the magnetic-support scenario have migrated to
the gravoturbulent one, despite it not being clear precisely
how they would fit into that scenario. Chief amongst these is the
notion that cores undergo a stage of quasistatic evolution, especially
during their prestellar stage. During this stage, the cores are
believed to be Jeans-stable, and confined by external pressure
\citep[e.g.,] [] {bert92, Lada+08}. This notion has been reinforced by the
frequent observation that cores exhibit Bonnor-Ebert-like (BE-like) column 
density profiles \citep[e.g] [] {alves01}, since BE spheres are known
solutions of the hydrostatic Lane-Emden equation. Proposals have been
made that, during this stage, the cores accrete material from their
surroundings, until they become Jeans-unstable and begin to collapse
dynamically \citep[e.g.,] [] {Simpson11}.

However, it is not clear how the cores could achieve a quasistatic
configuration if they are produced by dynamic compressions in a
supersonically turbulent medium, as the gravoturbulent scenario
proposes, and several complications arise. First, a Jeans-stable,
pressure-confined configuration requires the presence of a tenuous
confining medium {\it for the dense cores} -- i.e., at the sub-parsec
scale---implying that MCs should be two-phase media even down to these
scales. This is because equilibrium configurations in single-phase media
are generally unstable, and therefore not expected to be realized in a
supersonically turbulent medium \citep{VS+05}.  Second, it is not at all
clear why a hydrostatic configuration would arise first, and then
continue to accrete quasistatically from its surrounding medium if it
was initially formed by a turbulent compression that was dynamic to
begin with. In this case, evolving configurations are formed that
accrete through shocks. Before the shock-confined structure becomes
Jeans-unstable, it expands, and once it does become unstable, it begins
to collapse \citep{Gomez+07}, and never undergoes a quasistatic stage.
It is often suggested that perhaps the turbulent cascade feeds turbulence 
within the cores that supports them until it is dissipated, at which time 
they become unstable and collapse \citep[e.g.,] [] {BT07, keto15}. However, 
simulations of collapsing turbulent clumps never show an intermediate stable 
stage, after which the collapse would resume at the local scale. Instead, 
the cores form and either disperse or proceed directly to collapse 
\citep[e.g.,] [] {VS+98, VS+05, VS+17, RG12, Offner+14, Murray+17}.

A possible resolution to these problems is offered by the recent proposal 
of global, hierarchical gravitational collapse of MCs \citep[] [hereafter 
Paper I] {BH04, HB07, vaz07, VS+09, HH08, rn15}. In this scenario, MCs are 
born turbulent because of the combined action of several instabilities during 
their assembly stage \citep{KI02, Heitsch+05, vaz06}, but are also strongly
Jeans-unstable (i.e., containing {\it many} Jeans masses), because the
gas entering them suffers a phase transition from the warm-diffuse
atomic phase to the dense-cold phase, thereby increasing their density
and decreasing their temperature by a factor of $\sim10^2$, causing a
reduction of their Jeans mass ($\propto \rho^{-1/2} T^{3/2}$) by a
factor of $\sim10^4$ \citep{GV14}. Therefore, the clouds engage in
global gravitational contraction. As they do, the average Jeans mass in
the cloud decreases, because of the increasing mean density (at roughly
constant temperature), and then the small-scale density fluctuations,
induced by the turbulence, can undergo a collapse of their own when
their mass surpasses the average Jeans mass in the cloud. This process
then consists of collapses within collapses \citep{Heitsch+08,
VS+09}, and is essentially the same as Hoyle fragmentation
\citep{Hoyle53}, except that the density fluctuations are nonlinear, due
to the moderate turbulence induced into the cloud during
formation. The nonlinearity of the fluctuations and the
multi-Jeans-mass nature of the clouds eliminates
early concerns about this mechanism \citep{Tohline80}.

In Paper I we presented a numerical simulation of the prestellar
collapse of a local, near-Jeans-mass fluctuation embedded in a nearly
uniform, multi-Jeans mass medium, with spherical symmetry. This
simulation aimed at being an idealized model of the onset of
collapse of a fluctuation located in the outskirts of a much larger-scale 
unstable object, so that the large-scale collapse center was located outside
of the numerical box. The simulation was similar to the classic
calculations of \citet{lar69}, except for the additional ingredient of
being embedded in a uniform, strongly Jeans unstable background. Note 
that the initial fluctuation was very moderate, of only a 50\% enhancement 
above the background level.

This simulation produced a number of interesting features. First, it
developed a BE-like density profile (with a flat central
region and a power-law envelope), but {\it it was never in equilibrium},
because the fluctuation only provided a focusing center to the
collapsing tendency of the uniform background. So, even during the
stages when the ratio of the central to the boundary density (with the
boundary defined as the radius where the core merges into the
background) was less than the critical value for BE sphere
instability, the fluctuation was in the process of collapse. This
naturally explains the growth of cores when they appear to be
Jeans-stable and pressure confined. Second, the core developed an
``outside-in'' velocity profile \citep[see also] [] {WS85, GO11},
whereby the central parts of the core have a velocity that increases
{\it linearly} with radius, and the outer parts have a uniform velocity,
out to where the collapsing region extends. Third, the core, with its
boundary defined as explained above, traces the locus of both low- and
high-mass cores in a diagram of $\Mc/\MBE$ {\it versus} $\Mc$, first
proposed by \citet{Lada+08}, where $\Mc$ is the core's mass and $\MBE$
is its BE mass, even in the region occupied by ``stable''
cores. This result suggested that both low- and high-mass cores are
essentially Jeans-mass fluctuations undergoing collapse in a multi-Jeans
mass medium.

\subsection{Infall speed determination from molecular line profiles}
\label{sec:infall_determ}

However, one main caveat to this picture remained: that the simulated
core developed supersonic infall speeds shortly before it forms a
singularity (a protostar), in contradiction with the generally-accepted
notion that low-mass starless cores, in particular, exhibit subsonic,
not supersonic, infall speeds, as determined from their molecular-line
emission
\citep[e.g.,] [] {Lee01}. Indeed, moderately optically thick molecular
line transitions of non-homologously collapsing cores are known to
produce a self-reversed (i.e., with a self-absorption central dip)
profile with a blue excess
\citep[e.g.,] [] {snell77,zhou92,Evans99} or, if the line is narrow, a
single blue peak with a red shoulder. However, as dicussed by
\citet{Evans99}, blue-excess profiles can be caused by several other
mechanisms \citep[see also][]{leu77}, and so he advised that, to be a 
plausible candidate for collapse, a core must exhibit both (i) the 
self-reversed blue-skewed profile in a suitable optically-thick transition 
and (ii) a gaussian-like profile in an optically-thin line peaking at the 
self-absorption dip of (i). For (ii), the strength and skewness must peak 
at the central source. In this way, the two peaks of the blue-skewed profile 
should not be caused by clumps in an outflow.

To infer the infall speed of an observed core from a blue-skewed 
line profile, it is necessary to make an assumption on the form of
the radial velocity profile in the core. One of the earliest assumptions
made \citep[e.g.,] [] {snell77, zhou93} was that of an inside-out
collapse, resulting from the collapse of an initially static singular
isothermal sphere \citep[SIS;] [] {Shu77}.  However, this solution of
the spherical, isothermal collapse has been deemed unrealistic
\citep{Whitworth+96} because the SIS is an {\it unstable} equilibrium,
that is not expected to occur in general, let alone as a consequence of
a dynamic turbulent compression. Therefore, other assumptions for the
radial velocity profile have been made, including a simple two-layer
model \citep[e.g.,] [] {Myers96, Lee01}, and an initially unstable BE
(or BE-like) sphere \citep[e.g.,] [] {whit01, keto15}. 

Instead, the calculation presented in Paper I allowed the radial density and
velocity profiles to develop self-consistently from the instant the core was 
a minor perturbation in the background flow. This core could collapse even 
when its amplitude was only 50\% above the mean density because the entire 
simulation was gravitationally unstable, thus essentially only providing a focusing
center for the collapse of the whole region.

As explained above, this setup allowed a natural definition of the
instantaneous radial extent of the core, namely the radius at which
the perturbation merged into the uniform background. This boundary 
definition is inspired by observations, since observed cores are usually 
either defined down to the noise level or to where they merge with the 
background \citep[e.g.,] [] {and14}. This kind of definition, however, does
not imply that the cores actually end, or have a sharp density
discontinuity at their operationally-defined boundaries. Instead,
cores are known to be, in general, part of an extended continuum,
which consists of their parent molecular clouds. With this definition,
the numerical core grew both in mass and radius over
time, accreting material from the surrounding cloud,
and therefore does not coincide with any of the numerical model setups
used previously \citep[although see] [for a related
experiment] {MS13}, despite this being a more realistic situation for a 
dense core than, for example, a tenuous confining medium for the core 
\citep[e.g.,] [] {Ebert55, Bonnor56, Hunter77, Mousch76, KC10, keto15}.

Nevertheless, one remaining caveat of the simulation presented in
Paper I was that the infall speeds developed by the model become
supersonic during the latest stages of the prestellar evolution, in 
apparent contradiction with the observed subsonic speeds typically 
measured in low-mass cores \citep{Lee01}. This apparent contradiction in 
fact led \citet{MS13} to conclude that a similar numerical experiment, 
with a fixed accretion rate from its environment, was not realistic. 

In the present paper, we choose instead to investigate whether the
apparently subsonic observed velocities may in fact be an artifact of an
unrealistic assumption for the underlying core radial velocity 
profile. This is because in our core the collapse proceeds in an
outside-in fashion, with the maximum infall speed occurring at the
envelope of the core, not its center, as in the inside-out collapse of
\citet{Shu77}.  Since line profiles are essentially density-weighted 
line-of-sight (LOS) velocity histograms modified by absorption, 
having the maximum speeds at radii where the density is decreasing
suggests that these high infall speeds are down-weighted by the declining 
density in the envelope.

Various forms of a similar radial velocity profile, where the peak is
outside the core center, already exist in the
literature. \citet{SHLee07} found the contraction speed for L694-2 to
be small at the envelope, rising to a maximum of 0.28~$\kms$ at a
distance of $\sim$~0.08~pc from the center and decreasing inwards
thereafter. Analogously, with a similar velocity profile, they found a
peak infall speed of 0.2~$\kms$ at 0.095~pc from the center for
L1197. Such velocity profiles more satisfactorily match the broader
linewidths observed in both cores from their single-dish HCN $J=1-0$
data.

One of the classic indicators of collapse is the presence of
a {\it blue-skewed} line profile, which is defined as a bimodal profile 
in which the blue peak intensity ($T_{\rm b}$) exceeds the red peak 
intensity ($T_{\rm r}$), i.e. $T_{\rm b}/T_{\rm r}>1$. Numerous methods 
have been devised in the literature to interpret these profiles and derive 
infall velocities from them \citep[e.g.,] [] {Anglada+87, Anglada+91, 
zhou92, zhou93, Myers96, Evans99, dev05}. For example, the simple two-layer 
analytical model of \citet{Myers96} predicts that
$T_{\rm b}/T_{\rm r}$ increases with increasing $v_{\rm in}$. This ratio 
appears extreme among a number of starless and protostellar cores. The 
low-mass starless core B133 displays a value of $3.14\pm0.45$ \citep[\hcop $J=3-2$,][]{greg00}, 
while a value of $3.40\pm0.25$ is observed towards the high mass protostar 
associated with NGC 7538 \citep[\hcop $J=1-0$,][]{sun09}. Such values in early 
collapsing cores are too large to be explained by current collapse theories 
\citep{greg00}. \citet{Mard98} found the greatest $T_{\rm b}/T_{\rm r}$--ratios 
for a model in which the entire cloud is collapsing.

Furthermore, observations of line asymmetry in CS and \hcop\ lines 
far from the core center combined with these extreme $T_{\rm b}/T_{\rm r}$--values 
strengthen the case for an uninhibited gravitational infall. 

To investigate the relationship between the actual infall
speeds and those inferred from the line profile using 
typical line-modeling techniques, we present, in this
paper, synthetic spectral line observations of a numerical simulation of the
collapse of a spherical core in the context of the global hierarchical
gravitational collapse of a cloud; i.e., the collapsing core is
embedded in a globally unstable background medium.
To this end, we briefly discuss the numerical simulation, the bases for
selecting the lines and a prescription for the synthetic spectra 
(\S\ref{sec:input_model}). We then describe the various algorithms we apply to
catergorize the infall signature (\S\ref{sec:interp_tools}) and subsequently
infer infall speeds from the blue-excess synthetic lines. Finally, we 
present the results of this work (\S\ref{sec:results}), a discussion of
these results in the context of existing observational literature 
(\S\ref{sec:disc}) and our conclusions (\S\ref{sec:concls}).

\section{The Input Model} \label{sec:input_model}

\subsection{The numerical simulation} \label{sec:num_sim}

To produce synthetic spectra for a collapsing core, we consider 
a numerical simulation, presented in Paper I, of a collapsing spherically 
symmetric core inside a gravitationally unstable uniform background. As
discussed in Paper I, this configuration aims to represent the onset of
collapse of a density fluctuation, embedded in a large-scale collapse
flow focussing on a distant point in space, once it becomes locally
unstable. \citet{Longmore+14} have referred to this flow regime as a
``conveyor belt'' flow, and identified it as operating in the Central
Molecular Zone of the Galaxy. \citet{GV14} have also observed this type
of flow in simulations of cloud evolution where filaments form that feed
a massive hub, and local collapses occur within the filaments.

Due to limitations of the spectral code used for the simulation, and the
non-inclusion of sink particles, the collapse was followed only over the
prestellar stages of its evolution. Nevertheless, this allows us to
specifically investigate the generation of the initial conditions of
star formation.

In the simulation, the gas was isothermal, and initially at rest, with a
uniform density, $n = 10^4\, \pcc$, and a kinetic temperature, $T_{\rm
k} = 11.4$ K, implying an isothermal sound speed of $\cs = 0.2\,
\kms$. At this density and temperature, the Jeans length was $\LJ \equiv
(\pi \cs^2/G \rho)^{1/2} \approx 0.22$ pc, where a mean particle weight
$\mu = 2.36$ has been assumed. The numerical box had a size $\Lb =
\sqrt{10}\, \LJ \approx 0.71$ pc per side, implying that it had a total
mass $M \approx 207\, \Msun \approx 31.6\, \MJ$, where $\MJ$ is the
Jeans mass. The resolution was $512$ cells per dimension, implying a cell 
size of $1.4 \times 10^{-3}$ pc. For the generation of the synthetic spectra, 
we consider only the central half-length sub-box of the simulation, of 
volume $0.047\, {\rm pc}^3\, (=(0.71\, {\rm pc}/2)^3$), centered on the 
highest density volume pixel (voxel). 

On top of the uniform density field, a gaussian fluctuation of amplitude
50\% and standard deviation equal to $\LJ/2\, $ was added, so that the
density at the fluctuation peak was $1.5 \times 10^4\, \pcc$. The
fluctuation contains slightly more than one Jeans mass at 
the mean simulation density, although it would be undetectable by any standard 
observation because of the low contrast with the background. The mass increase 
caused by the fluctuation is so small that the mean density in the simulation
does not vary appreciably from the background value of $10^4\, \pcc$.

Because the entire numerical box is Jeans-unstable, gravitational
collapse ensues, focused towards the perturbation 
(which acts as a seed). The simulation terminates immediately before a 
singularity is produced, at $t \approx 2.14 \tff$, where $\tff =
\sqrt{3\pi/32G\rho} \approx 0.33$ Myr is the free-fall time at the mean
density. The actual collapse time is longer than the free-fall time because 
during the initial stages, the gravitational energy of the fluctuation is 
almost balanced by the internal pressure gradient, as the fluctuation
contains only a little over one Jeans mass \citep{lar69}. As the
collapse proceeds, gravity becomes increasingly dominant, and the
collapse rate asymptotically approaches the free-fall rate.

As described in Paper I, during its prestellar evolution,
the core develops a density structure that resembles a BE sphere, but
with a finite infall speed over a range of radii, rather than being
hydrostatic. The core has an {\it inner} region where the density is 
nearly uniform and the infall speed increases nearly linearly with 
radius, and an {\it outer envelope}, where the density approaches an 
$r^{-2}$ profile and the infall speed is roughly uniform, in agreement 
with the ``Band 0'' solution of \citet{WS85}. The transition between 
the inner flat-density region and the outer envelope occurs at a radius 
of the order of the Jeans length for the central density and temperature 
\citep{KC10}. 

However, the feature that distinguishes our simulation from other
simulations in which the core is artificially truncated at some
radius, is that, due to the presence of the uniform unstable
background, the density of our core does not continue to
self-similarly decrease as $r^{-2}$ at arbitrarily long distances, but
rather eventually reaches the background value, at which point the
core merges into the background, and the density remains flat beyond
this radius. This merging radius constitutes the boundary of our
core. As time progresses, the boundary moves outwards at a fraction of
the sound speed. For example, as indicated in Table 2 of Paper I, when
the core is defined as a 12.5\% enhancement over the background
density, its radius grows from 0.074 pc at $t=0$ to 0.19 pc at $t =
0.72$ Myr,\footnote{Note that there are typos in Table 2 of Paper I:
the core's radii are in units of 0.1 pc, not of 1 pc, as indicated,
and the times in Myr should be 0.23 and 0.72 instead of the indicated
0.73 and 2.27, respectively.} implying that the radius grows at a
speed $\sim 0.16\, \kms$, while the sound speed is $0.2\, \kms$. We
refer to this boundary as $\Rminifty(t)$,\footnote{We refer to this
boundary as $\Rminifty$ because in analytical and numerical studies of
collapse \citep[e.g.] [] {lar69, Hunter77, WS85, FC93} it is customary
to consider $t=0$ as the time of the formation of the
singularity, so that the prestellar evolution corresponds to negative
times. In this sense, all of the evolution of our prestellar core
corresponds to negative times in that convention, ending at $t
\rightarrow 0-$.} and corresponds to a rarefaction front which started
to propagate when the fluctuation started to collapse, and so it is
much larger than the standard position of the rarefaction front of the
classical inside-out solution of \citet{Shu77}, to which we refer as
$\Rshu$(t), which only starts to propagate at the time of the
formation of the protostar; i.e., at the ending time of our
simulation. We further discuss the implications of this behavior in
\S\ref{sec:collapse_models}.

Beyond $\Rminifty$, the infall speed decreases again, being zero at
the simulation boundary. Although the precise value of zero velocity
is an artifact of the periodic boundary condition, qualitatively the
decreasing nature of the infall speed is observed to occur even in
large-scale simulations where the local collapses are far removed from
the boundaries \citep[e.g.,] [] {GV14}. This is due to the
fact that, at long distances, the density tends to decrease even in
the extended cloud, and so the large-scale collapse is non-homologous,
with higher density inner regions infalling faster than outer,
lower-density ones.

As already mentioned above, we refer to the collapse regime of our
core as ``outside-in'', since the maimum infall speeds occur at a
finite radius from the center, of the order of the Jeans length of the
central density \citep{WS85, KC10}, and this radius of maximum
velocity {\it approaches} the center at a speed that eventually
becomes supersonic, but without developing a shock until the time of
singularity formation.

\begin{figure}
\centering

\subfigure[Linear-linear radial velocity profile]{
        \label{subfig:radialvel}
        \includegraphics[width=0.45\textwidth]{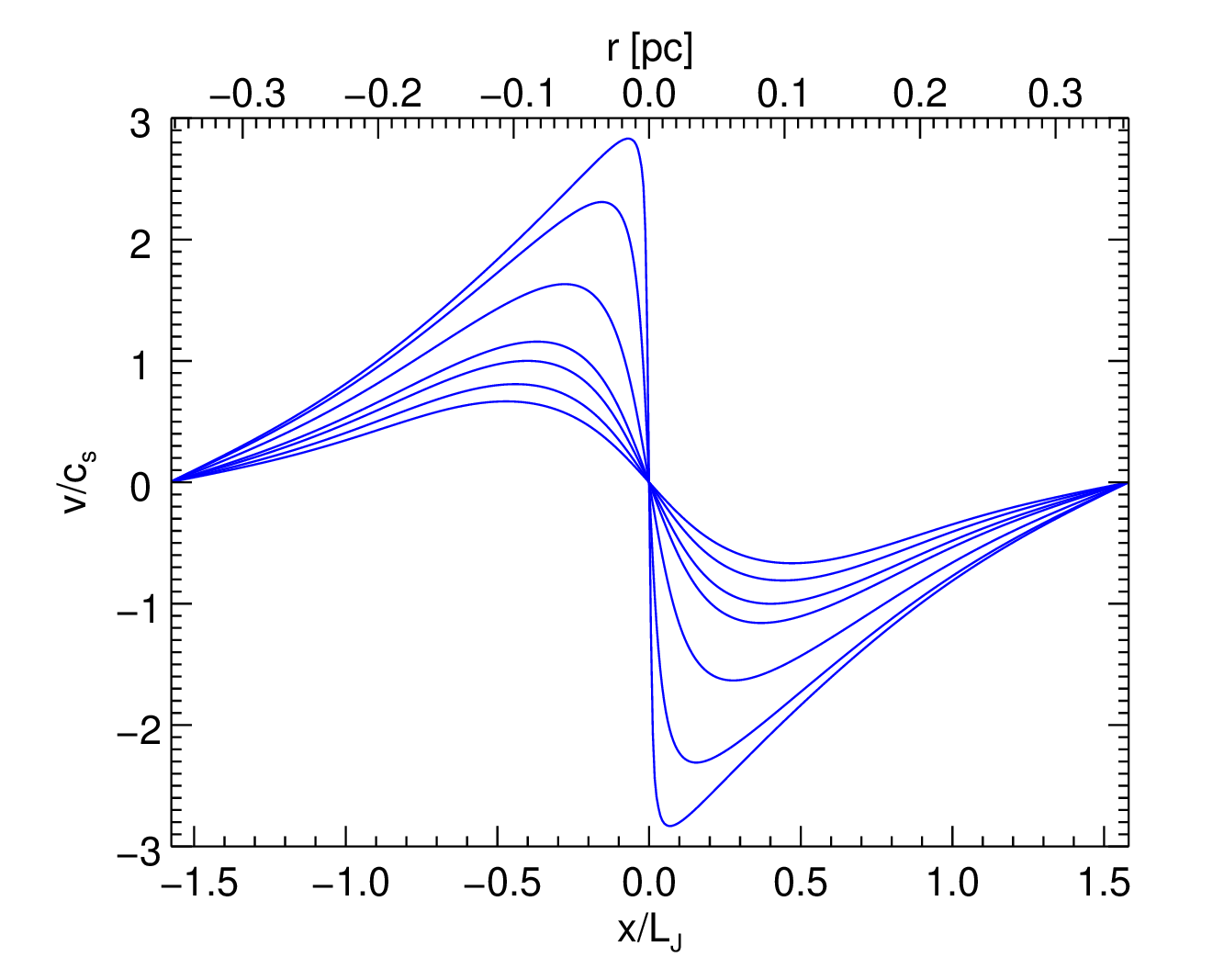}}

\subfigure[Log-linear radial density profile]{
        \label{subfig:radialdens}
        \includegraphics[width=0.47\textwidth]{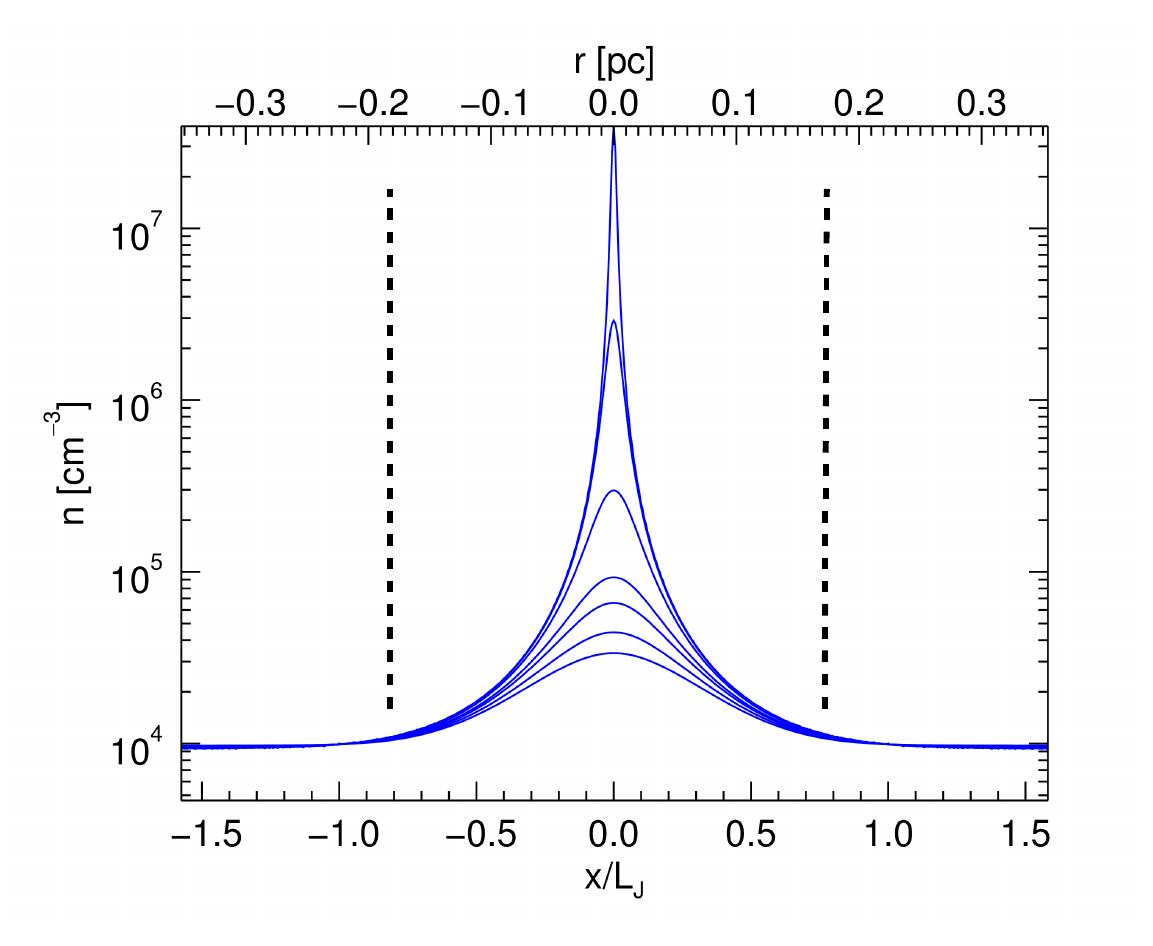}}
       
      \caption{Radial profiles for (a) velocity and (b) density 
	corresponding to the seven snapshots $49-65$ (inclusive) from 
	Table~\ref{tab:phystimestep}. The dashed vertical lines in each panel 
	demark the portion of the box considered in the radiative
	transfer analysis along each dimension (see \S\ref{sec:num_sim}). The 
	log-linear form of (b) is necessary due to the marked increase in 
	density between snapshots 64 and 65 prior to singularity formation. The 
	two panels taken together, along the radial dimension, highlight the 
	downweighting of higher velocity gas at the intermediate snapshots of 
	Table~\ref{tab:phystimestep} due to its location in the relatively lower 
	density envelope.}\label{fig:physrads}
\end{figure}

The fact that the infalling motions extend {\it beyond} the radius at which 
the core merges into the background \citep[Fig.\ \ref{subfig:radialvel}; see 
also] [] {MS13} implies the development of an accretion flow from the cloud onto
the core, a feature that cannot happen in simulations of collapse where
the core is artificially truncated at some finite radius, as is
customary \citep[e.g.] [] {lar69, Hunter77, FC93, keto15}. Although of
course our simulation also has a finite size, it is significantly larger
than the core's initial radius, allowing for the development of the
accretion from the cloud onto the core.

As the collapse progresses, the maximum infall speed along the radial
dimension--the nearly uniform speed in the envelope--increases with
time. We denote it by $\vmaxt$. This maximum infall speed becomes
transonic at $1.6 \tff$ (see Table~\ref{tab:phystimestep}). By the end
of the simulation, the maximum infall speed has reached $\sim
3\cs$, as is common in this kind of numerical simulation
\citep[e.g.,] [] {lar69}. 

A number of snapshots during the prestellar evolution are investigated.
Snapshot 49 was arbitrarily chosen as our earliest point along the 
evolutionary track of the collapsing core because at this time the density 
contrast between the core's maximum and the background is roughly a factor 
of 3. Note that at the final snapshot, number 65, the contrast is $\sim 2000$ 
(see fig.~\ref{subfig:radialdens}).

\begin{table}
\caption{Physical Parameters of Analysed Snapshots}\label{tab:phystimestep}
\begin{center}
\begin{threeparttable}
\begin{tabular}{cc|cc|cc}
\hline
No. & \multicolumn{2}{c} {\raisebox{1.5ex}{$t_{\rm evol}^a$ }} & \multicolumn{2}{c}{\raisebox{1.5ex}{$\vmaxt^b$}} & \raisebox{1.5ex}{$\Delta$x$^c$}\\
& \raisebox{1.0ex}{(Myr)} & \raisebox{1.0ex}{($\times t_{\rm ff}$)} & \raisebox{1.0ex}{($\times \cs$)} & \raisebox{1.0ex}{($\kms$)} & \raisebox{1.0ex}{(pc)}\\
\hline\hline
49 & 0.54122 & 1.4489 & 0.6668 & 0.1333 & 0.1051\\
52 & 0.57340 & 1.5453 & 0.8087 & 0.1616 & 0.0982\\
55 & 0.60688 & 1.6465 & 1.0005 & 0.1999 & 0.0899\\
57 & 0.62583 & 1.7095 & 1.1590 & 0.2316 & 0.0830\\
61 & 0.67301 & 2.0065 & 1.6327 & 0.3263 & 0.0622\\
64 & 0.70598 & 2.1048 & 2.3093 & 0.4165 & 0.0346\\
65 & 0.71702 & 2.1377 & 2.8218 & 0.5639 & 0.0152\\
\hline
\end{tabular}
\begin{tablenotes}
\item $^a$Time elaspsed since the beginning of the
  simulation, specified in terms of the free-fall time ($t_{\rm ff}$,
  \textit{see text}) and in mega-years (Myr).
\item $^b$Maximum infalling speed at each snapshot,
  specified in terms of the sound speed ($\cs$, see
    text) and in $\kms$.
\item $^c$Distance in parsecs (pc) of gas at velocity of $\vmaxt$
  from the simulation box center at each snapshot.
\end{tablenotes}
\end{threeparttable}
\end{center}
\end{table}

\subsection{Synthetic observations} \label{sec:synth}

We perform our synthetic observations using the accelerated (or
approximate) lambda iteration \citep[][]{rybick91} radiative transfer
code MOLLIE \citep{keto04, keto10}, which solves multi-level
non-LTE radiative transfer problems ensuring rapid convergence of the
radiation field and level populations. Assuming an initial input
continuum radiation field and LTE population, the code calculates the
new total radiation field and LTE populations. It then iterates until a
specified convergence criterion is achieved. At each radial point the
code generates (i) the level populations and (ii) the line source
function. We use a fairly stringent convergence criterion $\Delta
n_i/n_i\le10^{\rm-4}$, where $\Delta n_i$ represents the change in the
converged level population, $n_i$, between the $(i-1)^{\rm th}$ and
$i^{\rm th}$ iterations. The cosmic microwave background is the assumed
initial radiation field in the radiative transfer calculations.

The emergent intensity distributions are then convolved with an
appropriate synthetic telescope beam, so that the resulting spectra 
are comparable to observed line profiles from a given source observed 
with a specific telescope configuration. We assume that the telescope beam 
can be approximated by a Gaussian function, with a characteristic half-power 
beam width (HPBW). In typical observations of low-mass star-forming 
regions, the angular resolution (or beamwidth) of a single dish mm/sub-mm 
telescope is comparable to the angular size of nearby cores (e.g. Taurus-
Auriga or Perseus MC cores). For this work, we consider beamwidths that are
smaller than the simulated core dimension (up to $\sim20\%$ of the core
size).

\subsection{Considerations for the synthetic spectral data}
\label{sec:consid_syn_obs}

To produce synthetic spectral line data for our core, we first
need to choose suitable optically thin and moderately optically 
thick lines. For the latter, we choose to focus on transitions that have 
sufficiently high critical densities but that are excited throughout the core gas. 
\hcop\ is more opaque than either CS or H$_{\rm2}$CO and so its rotational transitions 
are better suited to revealing infall during the later stages when less material 
remains in the envelope \citep{greg00a}. The dipole moment of CS is 2.0~D, where D 
stands for $debyes$, while that of \hcop\ is 3.3~D implying greater sensitivity to 
the dynamics of the central core region. 

\hcop\ transitions are optically thick at the densities typical of 
molecular cores ($10^{3}-10^{5}$~cm$^{\rm-3}$) and so are useful for our
simulated snapshots where the background density is
10$^4$~cm$^{\rm-3}$. \hcop\ lines produce deeper self-absorptions than
CS lines of similar frequency, not solely because of the relatively higher
depletion of CS at higher densities towards the core centre, but because
\hcop\ predominates further out than CS and is therefore better able to
trace the higher velocity component of the infalling gas. A noticeable
feature in observational data is that \hcop\ lines exhibit higher
$T_{\rm b}$/$T_{\rm r}$--ratios than CS lines of similar
frequency. \citet{sun09} found, in their massive protostellar core
survey, that, on average, $T_{\rm b}/T_{\rm r}\sim2.2$ for HCO$^+$
$J=1-0$ and $\sim1.4$ for CS $J=2-1$, each with similar frequency. This
disparity is due to the excitation conditions required for each
transition and so each traces different spatial components of the core
gas along the LOS. For HCO$^+$, we consider a constant relative abundance of 
$3 \times 10^{-9}$ (to H$_2$).

For the thin tracer, we choose \nhp. This is a linear high-density
tracing species containing seven hyperfine components in its lower
$J=1-0$ rotational transition. Within this structure, the relatively
isolated hyperfine component, $J_{\rm F_1F}=1_{01}-0_{12}$, maintains a
near gaussian shape throughout the evolution of the simulated core. For
\nhp, we assume an abundance of 3$\times$10$^{-10}$ relative to H$_2$.

The ionized rotational collisional rate coefficients for collisions with 
$\rm H_{\rm2}$ from \citet{flow99} were used for the transitions of each of 
these species (see Table~\ref{tab:moltrans}). Since \hcop\ and \nhp\ have
identical molecular masses, this approximation is roughly warranted. The
modeled abundances for all considered transitions are typical of both
low mass dark clouds and intermediate mass core conditions
\citep{hoger97,kirk07}. 
 
Since the infall velocities at each step along the core's evolution are
comparable between adjacent snapshots (see Table~\ref{tab:phystimestep}) 
and the cloud is well radiatively coupled, the simple Sobolev large velocity gradient (LVG)
radiative transfer approximation is not applicable \citep{leu77}. Additionally, as 
microturbulent radiative transfer codes artificially add an extra turbulent contribution to
the velocity profile throughout the gas, hence creating broader line profiles \citep[such as those 
applied by ][]{zhou92}, we will not consider the microturbulent appoximation 
here \citep{Masu00}. A radiative transfer code that is capable of accounting for changes in physical 
parameters at the smallest scales, that is, non-LTE conditions, is more suitable.

\begin{table}
\caption{Modeled Molecular Transitions}
\begin{center}
\begin{threeparttable}
\begin{tabular}{lrr}
\toprule
Tracer$^{\dagger}$ & $\nu$ (GHz) & $n_{\rm cr}^*$ (cm$^{\rm-3}$)\\
\midrule
\hcop\\
    \hspace{1cm}$J=1-0$ & 89.18852 & 1.815$\times$10$^{5}$\\
    \hspace{1cm}$J=3-2$ & 267.55762 & 3.973$\times$10$^{6}$\\
N$_2$H$^+$\\
    \hspace{1cm}$J=1-0$ & 93.17370 & 1.549$\times$10$^{5}$\\
    \hspace{1cm}$J_{\rm F_1F}=1_{01}-0_{12}$ &
    93.17625$^{\ddagger}$ & 1.549$\times$10$^{5}$\\ 
\bottomrule
\end{tabular}
\begin{tablenotes}
\item $^*$Critical densities ($n_{\rm cr}=\nicefrac{A_{\rm
      ul}}{\sum_{\rm u} q_{\rm ul}}$) determined from the Leiden Atomic and Molecular
  Database (LAMDA).
\item $^{\dagger}$Each tracer has a molecular
  mass of 29 amu (1 amu~=~$1.660538 \times 10^{\rm-27}$ kg) resulting in an
  identical thermal velocity dispersion, $\sigma_{\rm
    th}=0.0572\, \kms$.
\item $^{\ddagger}$ The frequency of 93.17625~GHz actually corresponds to
  the location of three degenerate hyperfine lines: $J_{\rm
    F_1F}=1_{01}-0_{10}$, $1_{01}-0_{11}$, and $1_{01}-0_{12}$. The
  quantum number labelling of this line is routinely assigned to the
  latter on account of its higher relative intensity in the structure.
\end{tablenotes}
\end{threeparttable}
\label{tab:moltrans}
\end{center}
\end{table}

\begin{figure*}
\centering

\subfigure[Synthetic spectral positions for snapshot 52]{
        \label{subfig:ts52strip}
        \includegraphics[width=1.0\textwidth]{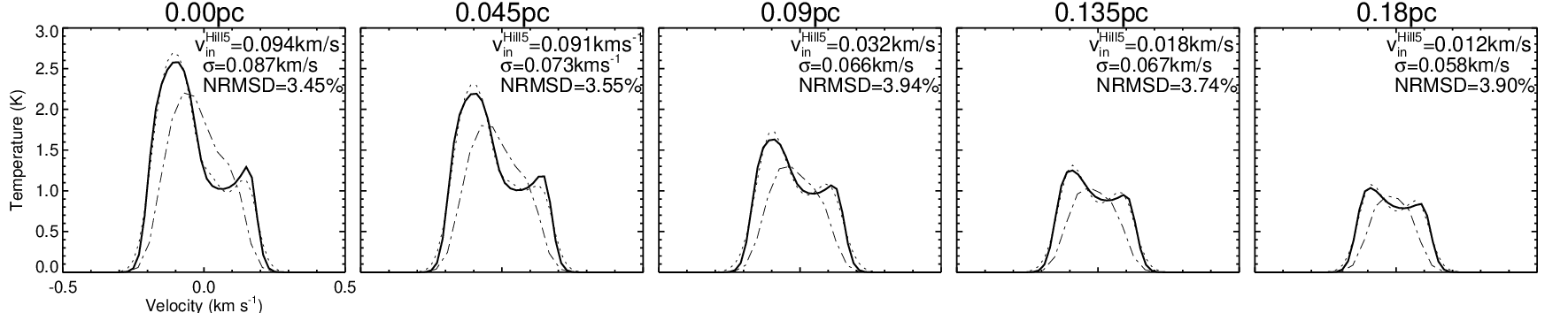}}

\subfigure[Synthetic spectral positions for snapshot 61]{
        \label{subfig:ts61strip}
        \includegraphics[width=1.0\textwidth]{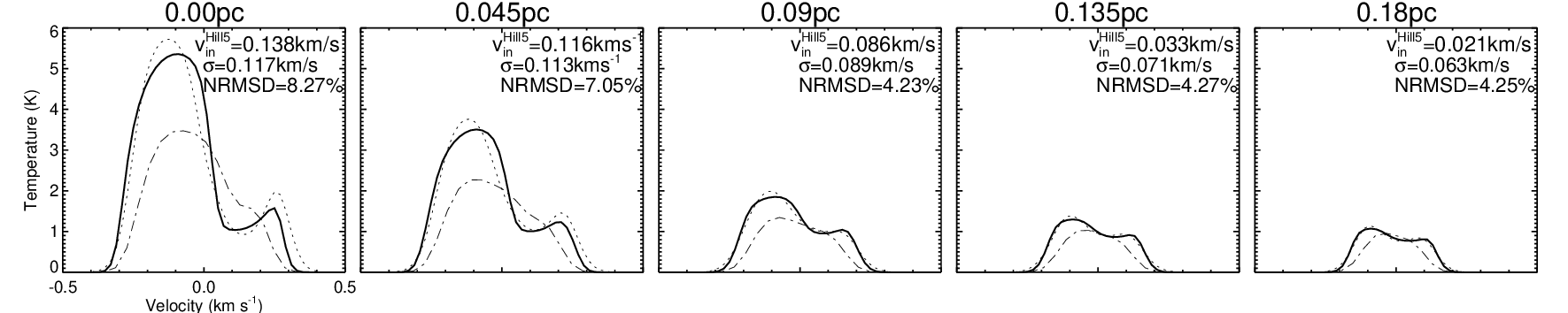}}
       
      \caption{Synthetic line spectra for (a) snapshot
        52 ({\it top row}) and (b) snapshot 61 ({\it bottom row}) at
        radial offsets $d$ (given in pc above each panel)
        from the core center. Each panel displays the \hcop\ $J=3-2$
        transition (solid line), the $\rm N_{\rm2}H^+$
        $J_{\rm F_1F}=1_{01}-0_{12}$ optically thin hyperfine component
        (dash-dotted line) and the best-fit Hill5 analytical model to the
        \hcop\ transition (dotted line) corresponding to the lowest 
        $NRMSD$-value. The beamwidth, $\theta_{\rm b}$, is 0.03~pc for 
        both spectral strips. The upper right-hand corner
        of each panel displays the Hill5 derived infall velocity, $v_{\rm in}^{Hill5}$, 
        the velocity dispersion of the $\rm N_{\rm2}H^+$ hyperfine line, 
        $\sigma$, and the $NRMSD$-value (see \S\ref{sec:goodfit}).
        }\label{fig:specpanels}
\end{figure*}

The resulting spectra from the radiative transfer runs are
Hanning-smoothed, which results in the number of channels being halved.
We initially tried a channel spacing of $0.01\, \kms$ for accurate
infall velocity determinations. The \nhp $J=1-0$ transition with its
quadrupolar hyperfine structure contained many more lines meaning a
larger spread in velocity. A coarser velocity resolution of $0.02\,
\kms$ was therefore favored.

We create synthetic spectra for \hcop\ rotational transitions. In
particular, we focus on two rotational transitions: $J=1-0$ \& $J=3-2$.
Higher rotational transitions are particularly useful in investigating
the velocity distribution in pre-stellar cores on account of their
higher critical densities ($n_{\rm cr}$, see Table~\ref{tab:moltrans}) than 
lower transitions of the same species since $n_{\rm cr}\propto\nu^3\mu^2$.

Three synthetic beamwidths, $\theta_{\rm b}$, are considered for the
synthetic spectra created in this work (in parsecs): 0.015, 0.03 and
0.06 pc, respectively. These correspond to 22$^{\prime\prime}$,
44$^{\prime\prime}$ and 88$^{\prime\prime}$ at the distance of the
Taurus-Auriga molecular complex ($\sim 140$ pc). The implemented
beamwidths in parsecs are independent of the perceived distance of the
simulated core. This is particularly useful in the comparison of our
findings with observational data.

Example strips of line profiles for snapshots 52 and 61 are displayed in 
fig.\ \ref{fig:specpanels} respectively at several radial positions on the
core of constant separation as projected on a virtual ``plane of the
sky''.

\section{Common spectral-line interpretation tools}
\label{sec:interp_tools} 

Observed line profiles are usually interpreted in terms of various
models of varying complexity, that focus on various features of the
profile to extract information about the infall regime. In this
section we review some of these techniques, so to later apply them
to our synthetic profiles, and thus see what would normally be inferred
from them. Thus, we can determine whether our numerical model and the
synthetic line profiles obtained from it using MOLLIE are consistent
with typical observed profiles, and how the standard interpretations
compare with the actual physical conditions in our core.

\subsection{Asymmetry Parameter: $\delta v$-Analysis} 
\label{sec:asym_par}

A first infall diagnostic for observed line profiles is the 
so-called {\it asymmetry parameter} \citep[$\delta v$,][]{Mard97}:
\begin{equation}
\delta v = \frac{V_{\rm thick}-V_{\rm thin}}{\Delta v_{\rm thin}},
\label{eq:asympar}
\end{equation}
where $V_{\rm thick}$ represents the peak velocity of the optically thick line    
(in our case, each of the \hcop\ $J=1-0$ and $3-2$ rotational lines). Likewise, 
$V_{\rm thin}$ represents the peak velocity of the optically thin line, and 
$\Delta v_{\rm thin}$ is its FWHM. A value $|\delta v| \geqslant 0.25$ indicates a 
strongly asymmetric line profile. Negative values imply blue-skewed and positive values 
imply red-skewed lines, respectively. 

The synthetic lines from the simulated core allow us to compute $\delta
v$ at each of the various evolutionary stages listed in Table
\ref{tab:phystimestep}, which can be directly compared to the values of this
parameter observed in real prestellar cores. Using \textsc{class} \citep{buiss94}, we
applied a global hyperfine fit to the N$_{\rm2}$H$^+$ $J=1-0$ hyperfine
components \citep[see Table 2 of][for degenerate and individual hyperfine
lines in this transition]{paga09} for all analysed synthetic spectral positions. 
By doing this, we were able to derive, explicitly, both 
$\Delta v_{\rm thin}$ and $V_{\rm thin}$ for use in eq.~\eqref{eq:asympar}.

The quantity $\Delta v_{\rm thin}$ is the FWHM of each of the N$_{\rm2}$H$^+$ $J=1-0$
hyperfine components (assumed constant) in each \textsc{class} fit and
$V_{\rm thin}$ is the relative shift in the peak velocity of the
N$_{\rm2}$H$^+$ $J=1-0$ transition, assumed to represent the systematic
motion of the cloud as a whole. This relative shift is brought about by
the infalling motions in our simulated core, but must be determined
with high precision so to obtain an accurate value for $\delta v$ from eq.\ (\ref{eq:asympar}). By using
high-accuracy laboratory frequencies for all hyperfine components \citep{Cas95} and
determining their relative shifts with respect to the
N$_{\rm2}$H$^+$ $J=1-0$ rotational transition frequency, we were able to
compute their corresponding rest velocity shifts relative to the central
``unshifted'' rotational line (at 0~km~s$^{-1}$). From the synthetic
N$_{\rm2}$H$^+$ $J=1-0$ data, we were then able to determine the relative
shifts of the individual hyperfine components from their rest relative
velocities. Taking the average of these relative shifts allowed a more
accurate determination of $V_{\rm thin}$. 

\begin{figure*}
\centering
\includegraphics[width=0.80\textwidth]{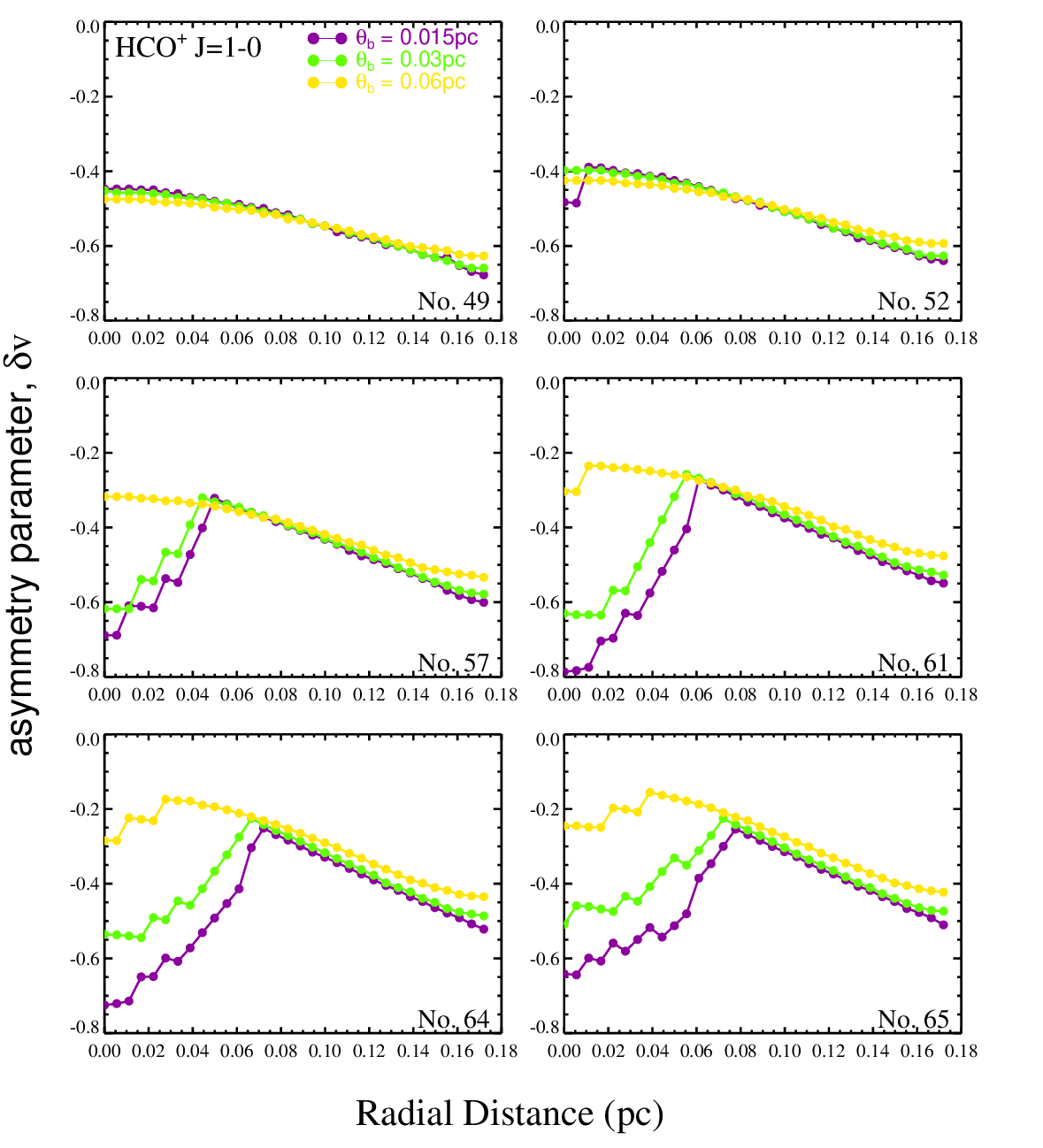}
\caption{Radial dependence of the asymmetry parameter, $\delta v$, for
the $J=1-0$ rotational transition. We include six snapshots in
ascending order approaching the end of the simulation. This transition
is affected by saturation, especially where $\theta_{\rm b}=0.015$~pc
($purple$) and 0.03~pc ($light~green$). For narrow beams, higher
density gas becomes saturated leading to a wider separation of the
peaks and hence larger values of $\left|\delta v\right|$. The extent
of this saturation diminishes for $\theta_{\rm b}=0.06$~pc ($yellow$),
where the derived values of $\left|\delta v\right|$ even out across
the synthetic spectral map resulting in snapshot $\delta v$-curves
resembling those for the HCO$^+$ $J=3-2$ transition (compare with
fig.~\ref{fig:asymplot2}).} 
\label{fig:asymplot1}
\end{figure*}

\begin{figure*}
\centering
\includegraphics[width=0.80\textwidth]{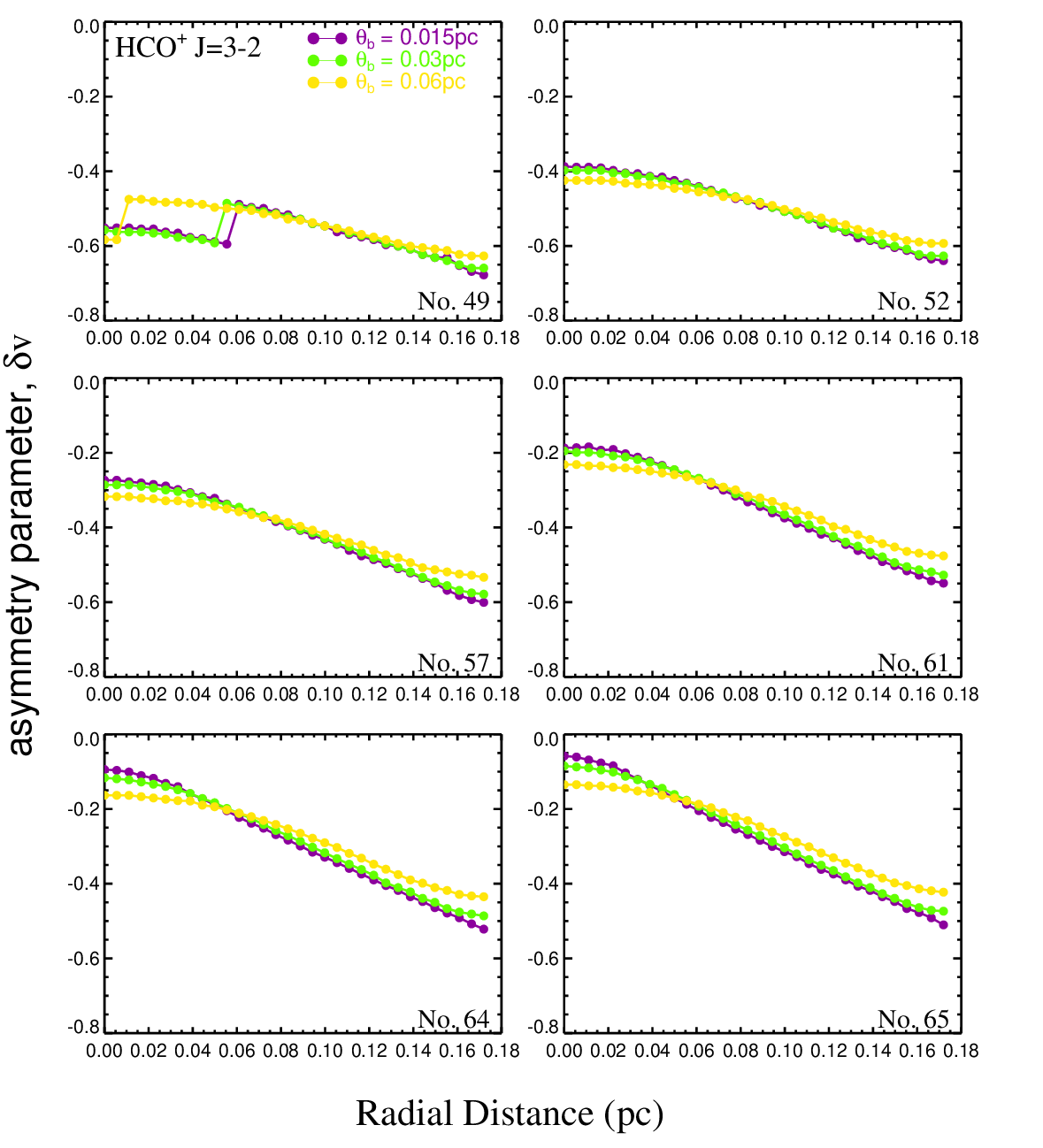}
\caption{Radial dependence of the asymmetry parameter, $\delta v$, for
the $J=3-2$ rotational transition. We include six snapshots in
ascending order approaching the end of the simulation. The $\delta
v$--values for this transitions are not overly affected by saturation 
and so one can infer the true radial variation of $\delta v$ and how that
depends separately on both $\theta_{\rm b}$ (0.015~pc ($purple$),
0.03~pc ($light~green$) and 0.06~pc ($yellow$)) and the evolution of
the core.} 
\label{fig:asymplot2}
\end{figure*}

The results from our $\delta v$--analysis are presented in 
figs.~\ref{fig:asymplot1} and \ref{fig:asymplot2}, for $J=1-0$ and $J=3-2$, respectively. 
The colored plots in each panel of these figures reflect a different value for    
$\theta_{\rm b}$ at which $\delta v$ was computed. We include a number of LOSs, of fixed separation (from the center),
towards the core, for snapshots 49-65 (inclusive) from Table~\ref{tab:phystimestep}.

\subsection{Hybrid Hill5 Analysis} \label{sec:hill5}

The general solution to the equation of transfer, assuming the optical
depth, $\tau$, increases away from the observer, is:

\begin{equation}
\Delta T_{\rm B} = T_{\rm i}e^{-\tau_{\rm0}} + \int_0^{\tau_{\rm0}} J(T_{\rm ex})e^{-\tau} d\tau
\label{eq:radsoln}
\end{equation}

\noindent
where $T_{\rm ex}$ is the excitation temperature of a region and
varies over the optical depth interval, (0, $\tau_{0}$), and
$T_{\rm i}$ is the incident specific intensity of radiation on that
region at $\tau_{\rm0}$, expressed in Kelvin. The Planck temperature,
$J(T)=(h\nu/k)/(\exp(h\nu/kT)-1)$ is also in Kelvin.

A number of simplifying assumptions on the nature of $J(T)$ exist whereby
the right-hand side of eq.~\eqref{eq:radsoln} can be expressed exactly.
The two-layer model \citep{Myers96} applies to an LOS in which two
regions of differing $T_{\rm ex}$ move towards one another with 
the near region having a lower $T_{\rm ex}$ ($T_{\rm f}$) than the far region 
($T_{\rm r}$). A constant $T_{\rm ex}$ is assumed along the LOS in this model. 
\citet{Lee01} used such a model to derive typical infall velocities of 
$\sim 0.1\, \kms$ for their starless core sample, comparable to the mean LOS linewidth,
$\sigma_{\rm los}$, in this sample. 

If $J$ in eq.~\eqref{eq:radsoln} is defined as a
linear function of $\tau$ (for all $\nu$) so that 
\begin{equation}
J(\tau)=J_{\rm1} + [(J_{\rm2}-J_{\rm1})/\tau_{\rm0}]\times\tau,
\label{eq:lintemp}
\end{equation}
\citep[][hereafter DM05]{dev05}, where $J_{\rm1}$ and $J_{\rm2}$ are constants, 
the eq.~\eqref{eq:radsoln} can be simplified as:
\begin{equation}
\Delta T_{\rm B} = T_{\rm i}e^{-\tau_{\rm0}} + (J_{\rm2}-J_{\rm1})\times\frac{1-e^{-\tau_{\rm0}}}{\tau_{\rm0}} + J_{\rm1} - J_{\rm2}e^{-\tau_{\rm0}}
\label{eq:stepjt}
\end{equation}
\noindent
where $\tau$ is a function of the Doppler velocity. DM05 defined their $Hill$ model 
using eqs.~\eqref{eq:lintemp} and \eqref{eq:stepjt}. This model includes a core with 
a peak $T_{\rm ex}=T_{\rm P}$ at the center (i.e. the maximum value of $J(\tau)$ in 
eq.~\eqref{eq:lintemp}), and $T_{\rm ex}=T_{\rm0}$ at both the near and far edges of 
the core (i.e. the minimum value of $J(\tau)$ for small $\tau$ in eq.~\eqref{eq:lintemp}). 
Since this model requires that $J(\tau)$ depends linearly on $\tau$ between these extrema, the 
excitation profile (i.e. $J(\tau)$) forms a ``hill''-profile (see fig. 1(b) of DM05).

The evolution of the two-layer model of \citet{Myers96} to the Hill model of
DM05 naturally occurred with our understanding that infall velocities 
are not simply determined from the separation of peaks in self-reversed line spectra. 
Since, although the velocity profile is identical in the two models, the provision of 
an inwardly-increasing $T_{\rm ex}$-profile in the latter enables the dependence
of the infall motion on the extent of the blue-red asymmetry, i.e. the 
$T_{\rm b}/T_{\rm r}$--ratio, to be accounted for. It is on this basis that the Hill 
models are more effective than the two-layer models at matching self-absorbed 
excitation profiles observed in starless cores.

To solve the equation of radiative transfer, the Hill $T_{\rm ex}$--profile is
split into two regions along the LOS: the first in the front part of the cloud, where 
$T_{\rm ex}$ increases along the LOS with optical depth $\tau_{\rm f}$ and the second in the rear 
part, where $T_{\rm ex}$ goes down along the LOS with optical depth $\tau_{\rm r}$, assuming a 
velocity dispersion of $\sigma$ in the entire cloud. The equation of transfer is then solved by
integrating along the LOS through each of the two regions to derive the
brightness temperature $\Delta T_{\rm B}(v)$, as a function of velocity.

The eight core parameters derived from the Hill model fit to the line profile
are $\tau_{\rm c}$ (the line center optical depth of the core), $\sigma$, $T_{0}$, $T_{\rm P}$, $v_{\rm lsr}$ 
(the core systematic velocity), $v_{\rm in}$, $v_{\rm E}$ and $\tau_{\rm E}$ (the respective 
optical depth and velocity of an (optional) external envelope, where $T_{\rm ex}=T_0$). 
The Hill5 is a variant of the Hill model where $T_{\rm0}$ is set to $T_{\rm b}$, the cosmic 
microwave background, with $v_{\rm E}=0$ and $\tau_{\rm E}=0$. According to DM05, 
beam smoothing does not adversely affect the accuracy of infall speeds obtained from this model.

We adopted a version of the Hill5 model, $Hill5hybrid$, that utilizes
a hybrid minimization algorithm consisting of the differential evolution (DE)
algorithm of \citet{storn1997} that initially separates the local minima
from the global minimum, for the combination of free parameters (discussed above) 
followed by Nelder-Mead simplex minimization \citep{nel65} to optimize 
the fit (DM05).

DE is an evolutionary algorithm that starts with a population of randomly
generated parameters. This parameter set is randomly modified during each
iteration until the global optimum solution is found. As the population evolves
at each iteration, it is referred to as a \textit{generation}.

The Hill5 routine takes six arguments. These are:

\begin{itemize}
\item rotational frequency in GHz (from Table~\ref{tab:moltrans}) 
\item $v_{\rm min}$: the furthest leftward non-zero velocity value
\item $v_{\rm max}$: the furthest rightward non-zero velocity value 
\item population in generation: the number of solutions to calculate each
  generation of the DE (used 300) 
\item generations per check: the number of generations to run before
  checking for convergence in the DE (used 300) 
\item checks to convergence: the number of checks to make before deciding
  the DE algorithm has converged (used 5) 
\end{itemize}

\begin{figure}
  \includegraphics[width=\linewidth]{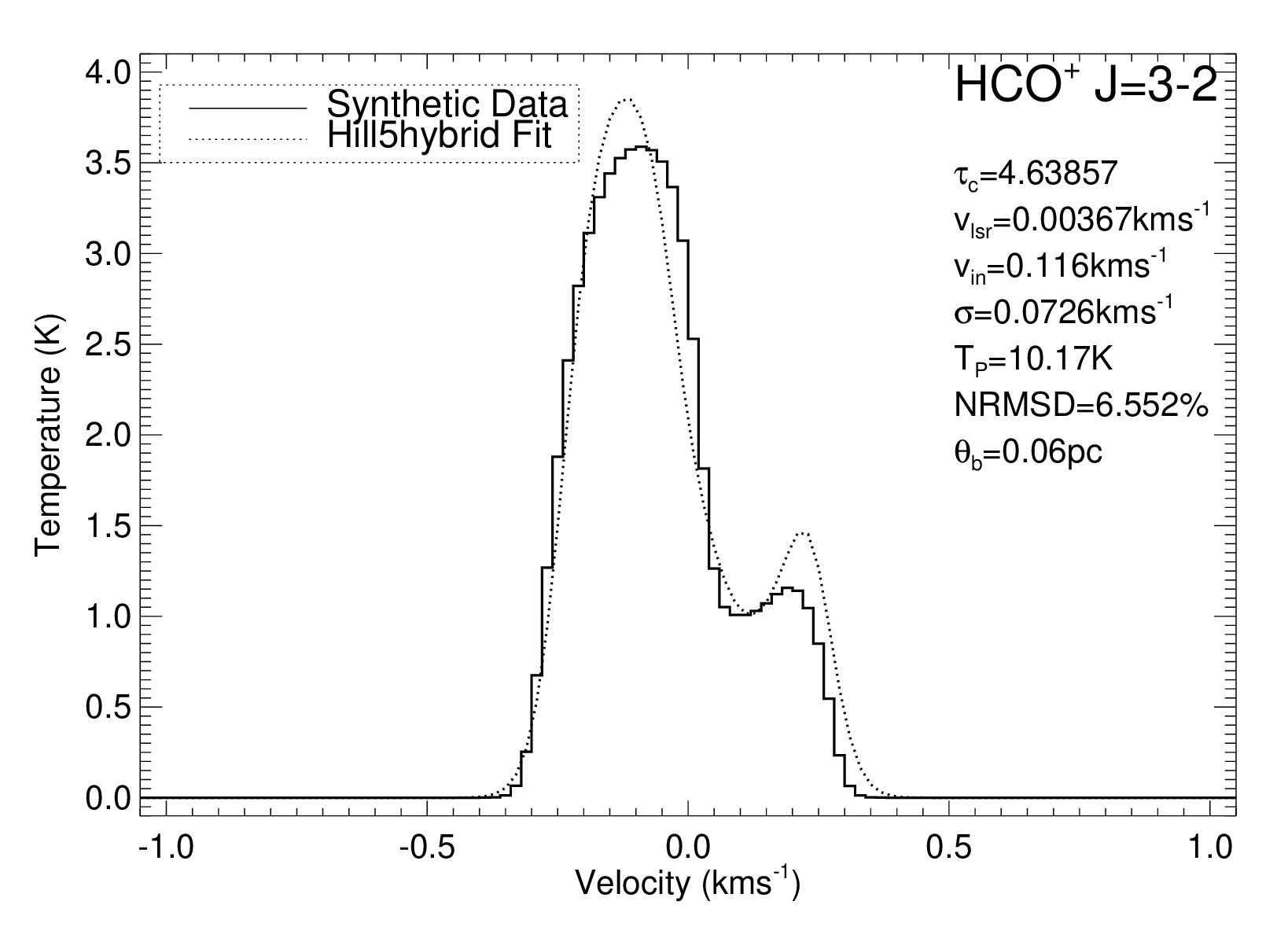}
  \caption{Synthetic spectrum for the \hcop\ $J=3-2$ transition at a
    beamwidth, $\theta_{\rm b}=0.06$~pc, at the core center for
    snapshot 61 (see Table~\ref{tab:phystimestep}). Overplotted (dotted-line) is 
    the Hill5 best-fit model to the synthetic data. Best-fit parameters are 
    displayed (\textit{see text}) together with the $NRMSD$-value 
    of the fit (see \S\ref{sec:goodfit}).}
  \label{fig:ts61_hco32}
\end{figure}

\begin{figure*}
\centering
\includegraphics[width=0.80\textwidth]{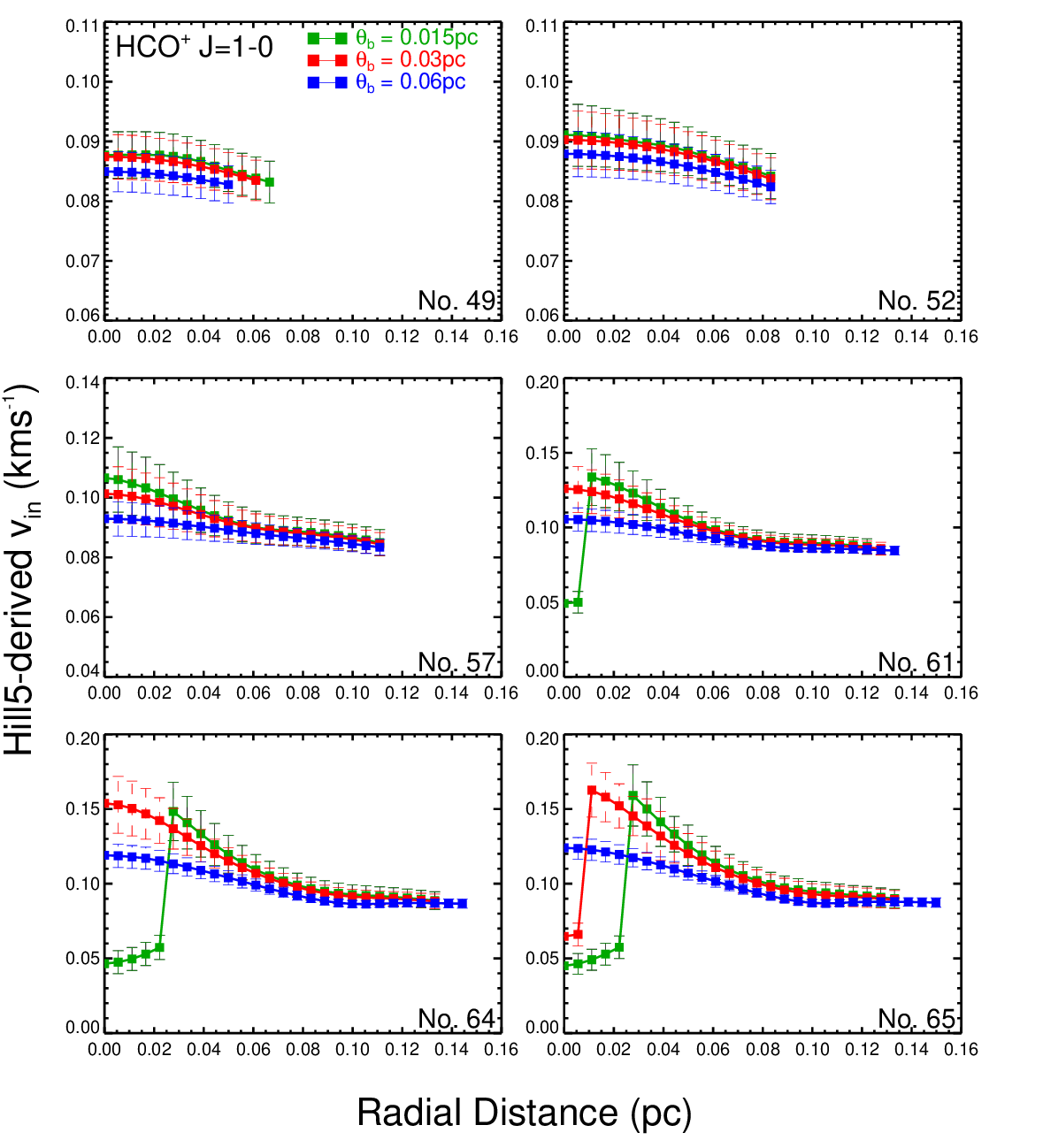}
\caption{The Hill5-derived infall velocities along multiple LOSs
through the core from its center to a position $\sim~$0.18~pc radially
offset from the center. We examine six snapshots in the core evolution
using the HCO$^+$ $J=1-0$ rotational transition and three values for
$\theta_{\rm b}$: 0.015~pc ($green$), 0.03~pc ($red$) and 0.06~pc
($blue$). Error--bars for the plotted values correspond to the
$NRMSD$--value for each Hill5--fitted spectrum. Artificial drops at the
center in several of the plots for the narrowest beamwidth correspond
to poorly fitted saturated spectra where the central density far
exceeds the transition $n_{\rm cr}$.} 
\label{fig:vin_plot}
\end{figure*}

\begin{figure*}
\centering
\includegraphics[width=0.80\textwidth]{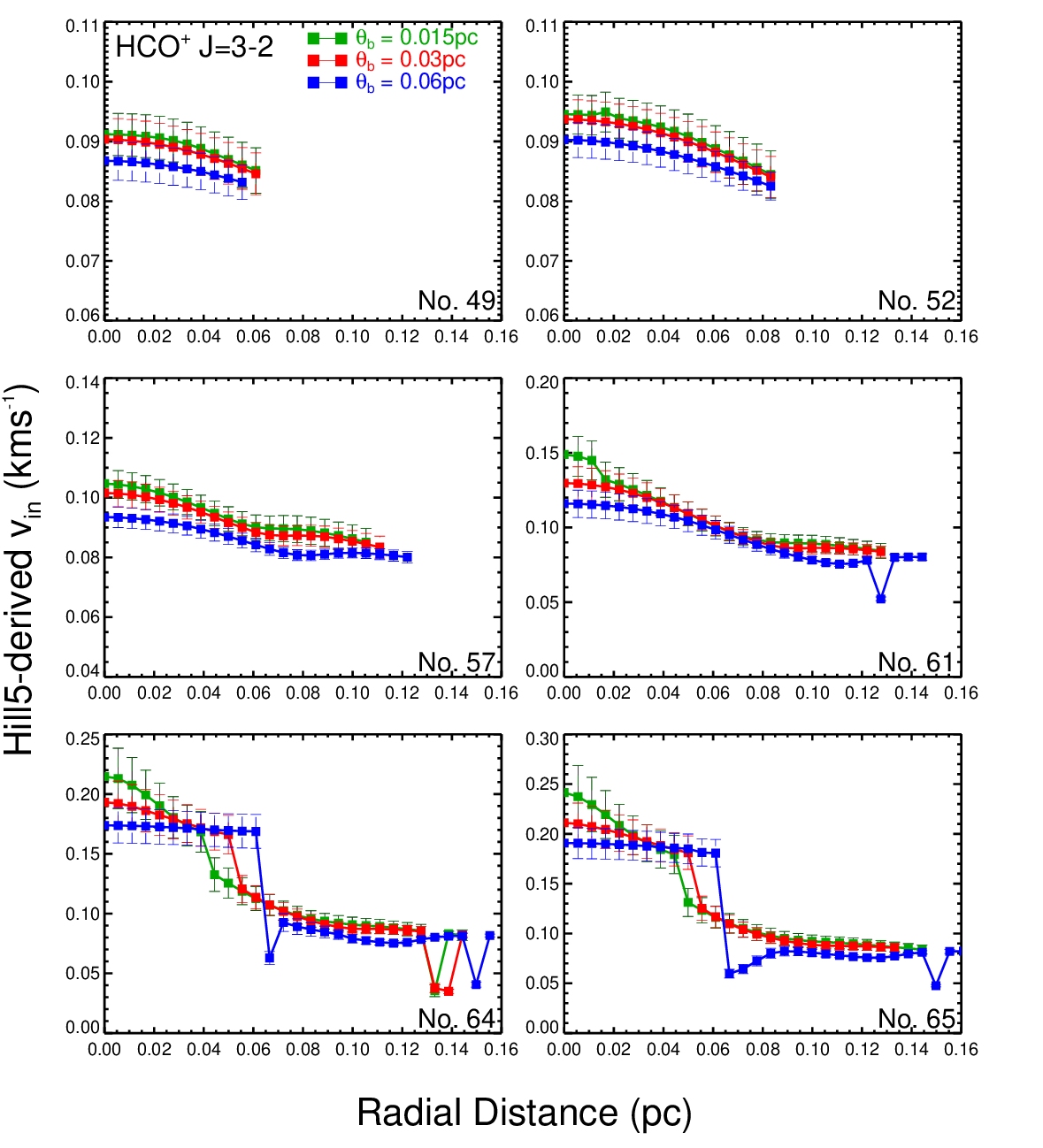}
\caption{The Hill5-derived infall velocities along multiple LOSs
through the core from its center to a position $\sim0.18$~pc radially
offset from the center. We examine six snapshots in the core evolution
using the HCO$^+$ $J=3-2$ rotational transition and three values for
$\theta_{\rm b}$: 0.015~pc ($green$), 0.03~pc ($red$) and 0.06~pc
($blue$). Error-bars for the plotted values correspond to the
$NRMSD$--value for each Hill5-fitted spectrum. Unlike the $J=1-0$
transition, $J=3-2$ does not suffer from the artificial drops for
small $\theta_{\rm b}$. However, for the later stages of the
simulation, where the velocity field increases steeply at the center
of the core, the larger beam of $\theta_{\rm b}=0.06$~pc samples the
field too coarsely, resulting in abrupt changes in its $v_{\rm
in}^{Hill5}$--radial plots (after snapshot 57) compared to
narrower $\theta_{\rm b}$--values.} 
\label{fig:vin_plot32}
\end{figure*}

Applying the model using the suggested parameters from DM05, the example fit in 
fig.~\ref{fig:ts61_hco32}, to a synthetic \hcop\ $J=3-2$ spectrum (for snapshot 61 at 
$\theta_{\rm b}=0.06$~pc), is derived. A detailed analysis of $v_{\rm in}$ 
towards the simulated core at a selection of the times listed in 
Table~\ref{tab:phystimestep} is given in \S\ref{sec:infall_speed} and displayed in 
figs.~\ref{fig:vin_plot} and \ref{fig:vin_plot32}. Unlike in fig.~\ref{fig:ts61_hco32}, 
this analysis excludes fit values for $\tau_{\rm c}$, $T_{\rm P}$ etc., since we only 
invoked the Hill5 model to determine the value of $v_{\rm in}$ for each synthetic spectral 
position during the core's evolution. In this way, the Hill5 model is used to exemplify the 
infall speed that known models would infer from the line spectra produced in this work. We 
do note, however, that $\tau_{\rm c}$ and $T_{\rm P}$ increase steadily with the density of 
the core gas, as expected.

\citet{Myers96} determined an analytical expression for $v_{\rm in}$ from self-absorbed optically thick lines in terms of observable line parameters. Assuming $v_{\rm in}\ll\sigma(2\ln\tau_0)^{\nicefrac{1}{2}}$, where $\sigma$ is the velocity dispersion of a concurrent optically thin line and $\tau_0$ is the line optical depth, they found \citep[equation 9 in][]{Myers96}:

\begin{equation}
v_{\rm in}^{an} \approx \frac{\sigma^2}{v_{\rm red}-v_{\rm blue}}\ln{\biggl[\frac{1+eT_{\rm BD}/T_{\rm D}}{1+eT_{\rm RD}/T_{\rm D}}\biggr]}.
\label{eq:analexp}
\end{equation}

\noindent
where $T_{\rm D}$ is the height of the self-absorption dip, $T_{\rm BD}$ ($T_{\rm RD}$) is the height of the blue (red) peak above the dip, and $v_{\rm blue}$ ($v_{\rm red}$) represents the position of the blue (red) peak. 

Tests, based on analytical models,\footnote{Modeled spectral profiles
formed with $\tau_0=5$ and 10, $T_{\rm k}=10$~K, $\sigma_{\rm
nth}=0.07~$km~s$^{-1}$ and $v_{\rm in}$ varying between 0 and
2$\sigma_{\rm nth}$.} for eq.~\eqref{eq:analexp} find that the 
analytically-derived value differs from the actual modeled $v_{\rm in}$ by 
up to 20\% \citep{Myers96}. Similarly, we determined the difference between the
Hill5-derived value ($v_{\rm in}^{Hill5}$) and the corresponding
analytical value ($v_{\rm in}^{an}$), from eq.~\eqref{eq:analexp}, at various 
positions from the core's center at the different snapshots in
Table~\ref{tab:phystimestep}. For no position does the difference exceed
20-25\%, with saturated positions at the latest snapshots falling into
the higher end of this range. As an example, by applying
eq.~\eqref{eq:analexp} to the spectrum in fig.~\ref{fig:ts61_hco32}, we
find $v_{\rm in}^{an}=0.095$~km~s$^{-1}$. From
fig.~\ref{fig:ts61_hco32}, $v_{\rm in}^{Hill5}=0.116$~km~s$^{-1}$, a
22\% difference from $v_{\rm in}^{an}$. This result is in excellent
agreement with the tests conducted by \citet{Myers96} given that those
tests restricted the infall velocity to values smaller than the velocity
dispersion.

We should note that, from our analysis, eq.~\eqref{eq:analexp} is only suitable for use with centrally-located, strongly emitting asymmetric line profiles in the latest stages of collapse that precede star formation. It fails to adequately match the infall velocity for positions widely separated from the core center and spectra associated with earlier stages of the collapse, i.e. spectra with low $T_{\rm b}/T_{\rm r}$--ratios.

\subsubsection{Goodness-Of-Fit}
\label{sec:goodfit}

To estimate the accuracy between the values predicted by the fit and the physically observed values for the core, we use the root mean square deviation ($RMSD$), defined as the square root of the mean squared error:

\begin{equation}
RMSD = \sqrt{\frac{\sum_n^{i=1}(y_{\rm o,i}-y_{\rm m,i})^2}{n}},
\label{eq:rmsdeq}
\end{equation}

\noindent
where, for this work, $y_{\rm m,i}$ is the $i$th estimated (or
modeled) data point and $y_{\rm o,i}$ is the corresponding synthetic
spectral data point, and the sum runs over the velocity channels
i.

A comparative measure of the accuracy of each spectral fit contained in this 
work is the normalized $RMSD$, or $NRMSD$, calculated by dividing the $RMSD$ 
by the range of the synthetic data values in each Hill5 model fit:

\begin{equation}
NRMSD = \frac{RMSD}{y_{\rm o,max}-y_{\rm o,min}}
\label{eq:nrmsdeq}
\end{equation}

\noindent
which we report as a percentage. Eq.~\eqref{eq:nrmsdeq} gives a suitable
relative, \textit{scale-invariant} measure \citep{hynd06} that can be
used to contrast the goodness-of-fit of the Hill5 model fits amongst all
modeled spectral positions at each of the studied snapshots in
Table~\ref{tab:phystimestep}. Relatively large $NRMSD$--values
($\gtrsim10$\%) indicate poor model fits where the underlying synthetic
spectral data are either too saturated (e.g.  $J=1-0$ data for snapshots
later than 61) or where the self-absorption dip 
(see fig.~\ref{fig:ts61_hco32} e.g., compare respective 
spectra in panels (a) and (b) from fig.~\ref{fig:specpanels}) is too shallow
(see \S\ref{sec:underestim} for further discussion).

\subsection{$T_{\rm b}/T_{\rm r}$ ratio}
\label{sec:tbtr_sec}

As mentioned in the Introduction, another indicator of infalling motions
is the degree of blue/red asymmetry observed in a sufficiently optically
thick line. The asymmetry is quantified by determining the ratio of the
blue to red peaks, or $T_{\rm b}/T_{\rm r}$ \citep[e.g.,
see][]{sun09}. A \textit{blue} profile requires $T_{\rm b}/T_{\rm
r}>1.0$, i.e. the blue peak is stronger than the red peak, while $T_{\rm
b}/T_{\rm r}<1.0$ implies a \textit{red} profile. Since all of our optically
thick synthetic spectral positions show two clear peaks, we computed
$T_{\rm b}/T_{\rm r}$ at selected positions in the spectral maps for the
selected snapshots in each of the \hcop\ $J=1-0$ and $J=3-2$ rotational
transitions. $T_{\rm b}$ and $T_{\rm r}$ are simply the values of the
blue and red peak emissions, in Kelvin, from each synthetic spectrum,
respectively. Spectra displaying large values of this ratios can appear in the form of a strong 
blue component with a relatively weaker shoulder. 
We analyse the optically thick synthetic spectra in terms of this ratio with the 
results presented in \S\ref{sec:temp_ratio_vs_r} and displayed in figs.~\ref{fig:tbtr_10} and 
\ref{fig:tbtr_32}.

\section{Results}
\label{sec:results}

\subsection{Comparison of line profile-derived speeds with actual
physical speeds} \label{sec:speed_compar}

From the radial plots of figs.~\ref{fig:vin_plot} and \ref{fig:vin_plot32}
(see also the example spectra in fig.\ \ref{fig:specpanels}), the inferred 
infall motions are seen to be subsonic in most cases, for both transitions,
in spite of the actual maximum infall speeds in the simulation being supersonic (see 
Table~\ref{tab:phystimestep}), up to Mach numbers of nearly 3. Our 
inferred infall speeds using the Hill5 method range from $\sim$~0.07 to 0.24 
$\kms$ at the center of the collapse for the studied snapshots, becoming 
marginally supersonic at the final snapshot only. Additionally, a simple 
inspection of the line profiles shows that the separation between the blue 
peak and the absorption minimum of the line lies between 0.15 and 0.2 $\kms$, thus
being perfectly consistent with reported infall speeds derived from
blue-skewed asymmetric profiles \citep[e.g.,] [] {Lee01,
Campbell+16}. {\it This implies that the standard modeling of
blue-asymmetry line profiles underestimates the infall speeds by factors
of up to} $\sim 3$--4.

\subsection{Radial variation of derived core parameters}

In this section, we examine the radial variation of the aforementioned
parameters at various timesteps in the simulation as the LOS is shifted
from the core center to a position $\sim$~0.18~pc from the center. We also 
examine the variability in the determined parameters with the value of 
$\theta_{\rm b}$, i.e. the effects of beam-averaging.

\subsubsection{Radial variation of $\delta v$}
\label{sec:mardon_rad}

In figs.~\ref{fig:asymplot1} and \ref{fig:asymplot2}, we plot the variation
of the asymmetry parameter, $\delta v$, with radial offset for HCO$^+$ $J=1-0$
and $J=3-2$, respectively. In these plots, we consider beamwidths of 0.015~pc, 
0.03~pc and 0.06~pc, respectively, for each of the analyzed snapshots. As line 
saturation occurs at all frequencies over a lineshape function, the radial 
profile of this parameter can be affected at small radial offsets where the
density and hence optical depth increase significantly towards the end of the
simulation (see \S\ref{sec:underestim} for an in-depth discussion of line 
saturation). 

\subsubsection{Radial variation of $v_{\rm in}^{Hill5}$}
\label{sec:infall_speed}

In fig.~\ref{fig:specpanels} we display the \hcop\ $J=3-2$ synthetic 
spectra from the LOS-projected core at 5 different radial positions, namely the 
central position as well as four neighboring positions at increasing distances from 
the core center (with a constant separation of 0.045 pc) for snapshots 52
(fig.~\ref{subfig:ts52strip}) and 61 (fig.~\ref{subfig:ts61strip}), respectively. 
For both sets of profiles, we take $\theta_{\rm b}=0.03$~pc. An overlaid best-fit 
hybrid Hill5 model as well as the isolated \nhp\ hyperfine component are also included 
for each position. To quantify the quality of each fit, we display the estimated infall 
velocity from the hybrid Hill5 model, $v_{\rm in}^{Hill5}$, the $NRMSD$--value of the
fit, and the value of $\sigma_{\rm los}$ from the isolated N$_2$H$^+$
hyperfine component.

The radial dependence of the Hill5-derived infall velocity ($v_{\rm
in}^{Hill5}$) for both the HCO$^+$ $J=1-0$ and $J=3-2$ 
transitions, is plotted for all snapshots in figs.~\ref{fig:vin_plot} and 
\ref{fig:vin_plot32}, respectively. The synthetic beamwidths, $\theta_{\rm b}$,
of 0.015~pc (green), 0.03~pc (red) and 0.06~pc (blue) are plotted
together for a number of radial offsets from the core center for each
snapshot in the two transitions. We have also converted the $NRMSD$--values
from each best-fit Hill5 model into an associated error of the fit and
plot these errors in the form of error bars for all analysed positions
in figs.~\ref{fig:vin_plot} and \ref{fig:vin_plot32}.

At this point we should note that there is an unresolvable
degeneracy between the optical depth in a line profile and either the
brightness ratio of its peaks ($T_{\rm b}$/$T_{\rm r}$) or their
velocity separation. That is to say, as the optical depth increases, the
separation of the peaks becomes unrelated to the actual velocity of the 
gas within the beam. However, as we clearly 
demonstrate in the radial plots for $v_{\rm in}^{Hill5}$ in figs.~\ref{fig:vin_plot}
and \ref{fig:vin_plot32}, very optically thick or saturated spectral
positions have high $NRMSD$-values in our analysis and,
though presented, are considered unreliable data.  

\subsubsection{Radial variation of $T_{\rm b}/T_{\rm r}$} 
\label{sec:temp_ratio_vs_r}

\begin{figure*}
\includegraphics[width=0.95\textwidth]{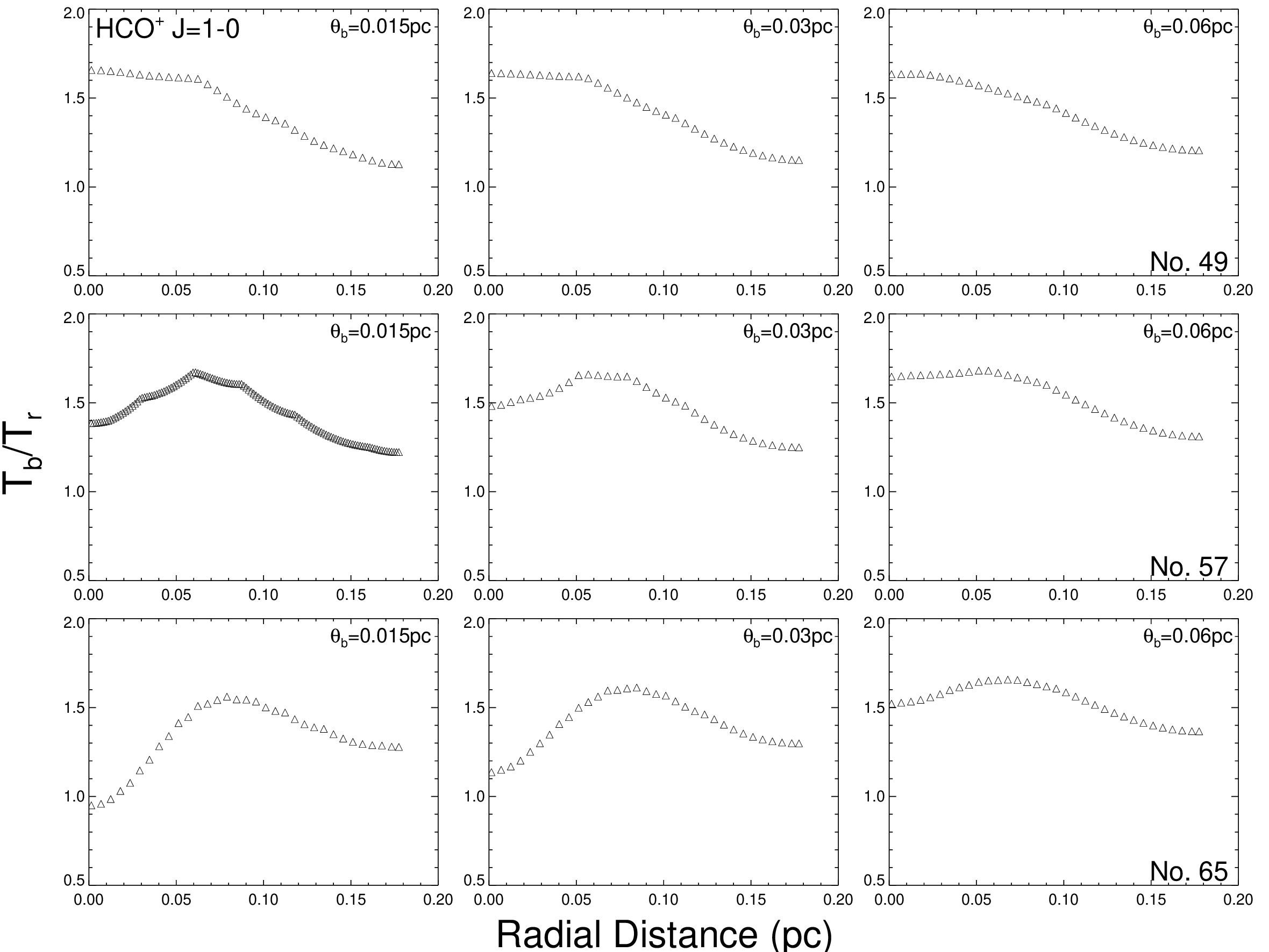}
\caption{Variation of $T_{\rm b}/T_{\rm r}$ (see \S\ref{sec:tbtr_sec}) 
  with radial distance from core center using \hcop\ $J=1-0$ for 
  snapshots 49 ($t=1.45\, \tff$, {\it top row}), 57 
  ($t=1.71\, \tff$, {\it middle row}) and 65 ($t=2.14\, \tff$, 
  {\it bottom row}) for the $J=1-0$ transition of \hcop\ for $\theta_{\rm b}= 0.015$ pc
  (\textit{left}), 0.03 pc (\textit{center}) and 0.06 pc
  (\textit{right}), respectively.}
\label{fig:tbtr_10}
\end{figure*}

\begin{figure*}
\centering
\includegraphics[width=0.95\textwidth]{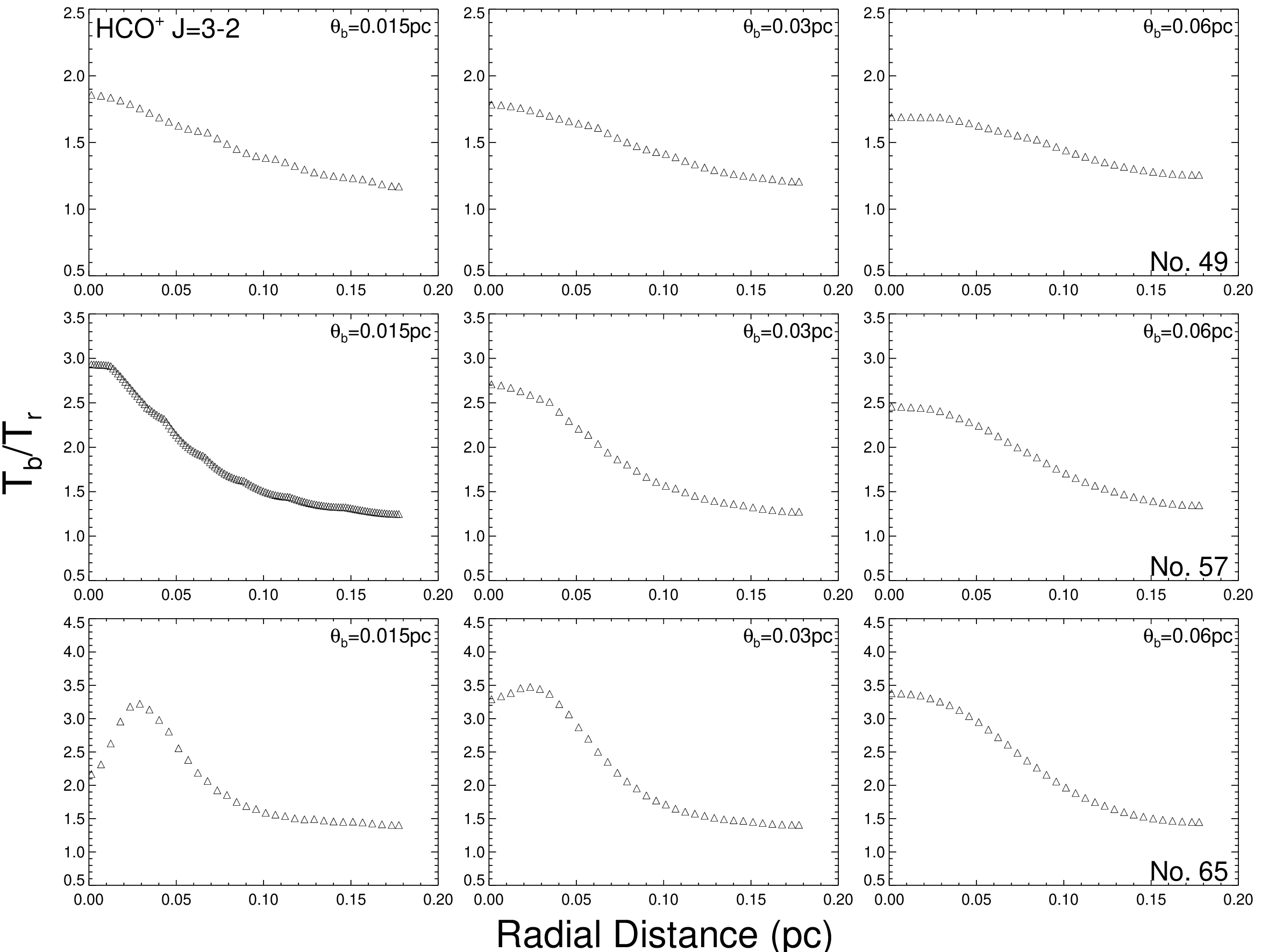}
\caption{Variation of $T_{\rm b}/T_{\rm r}$ (see \S\ref{sec:tbtr_sec}) 
  with radial distance from core center using \hcop\ $J=3-2$ for 
  snapshots 49 ($t=1.45\, \tff$, {\it top row}), 57 
  ($t=1.71\, \tff$, {\it middle row}) and 65 ($t=2.14\, \tff$, 
  {\it bottom row}) for the $J=1-0$ transition of \hcop\ for $\theta_{\rm b}= 0.015$ pc
  (\textit{left}), 0.03 pc (\textit{center}) and 0.06 pc
  (\textit{right}), respectively.}
\label{fig:tbtr_32}
\end{figure*}

The variation in $T_{\rm b}/T_{\rm r}$ with radial
distance from the core center for snapshots 49 ($t=1.45\, \tff$, {\it
top row}), 57 ($t=1.71\, \tff$, {\it middle row}) and 65 ($t=2.14\,
\tff$, {\it bottom row}) is presented in fig.~\ref{fig:tbtr_10} (\hcop\
$J=1-0$) and fig.~\ref{fig:tbtr_32} (\hcop\ $J=3-2$) for $\theta_{\rm
b}=0.015$ pc (\textit{left}), 0.03 pc (\textit{center}) and 0.06 pc
(\textit{right}), respectively. As an aside, for snapshot 57 at 
$\theta_{\rm b} = 0.015$ pc for each transition, we determined 
$T_{\rm b}/T_{\rm r}$ at the limiting spatial resolution of the input grid, 
i.e. at every voxel along a radial arm from the center to the edge of the 
$256^3$ subgrid. The derived variation, showing a superposed secondary 
spurious peak pattern for each transition, is indicative of the limiting spatial resolution of the 
simulation.

\noindent

\section{Discussion} \label{sec:disc}

\subsection{Comparison with previous observational studies}
\label{sec:underestim}

In \S \ref{sec:results} we have provided a testable framework for the 
hierarchical collapsing core model by means of analyzing its synthetic 
spectra in commonly observed species. Our findings can be directly compared 
with existing observed spectral maps to place a firmer understanding 
on the dynamics in star-forming cores.

Specifically, we have analyzed the synthetic spectra determined for a
number of snapshots within the last 25\% of the evolution of an isothermal,
spherically symmetric, hierarchically collapsing core, the last of which
corresponds to a time immediately prior to protostellar formation. 
With this simple model of a collapsing core in a gravitationally unstable 
envelope, we can reproduce the blue-skewed asymmetric line profile
indicative of collapse motions. We find, like \citet{zhou93}, that this
signature is visible also in higher excitation rotational lines.

On the other hand, \citet[][hereafter Z92]{zhou92} found an
extended line profile signature (or ``extended wing emission'') in a synthetic 
LP-model and concluded that the LP model was unrealistic, 
as this signature is generally not observed. 
However, it has been suggested that the microturbulent model used by Z92 
artificially broadened his synthetic line profiles \citep{Masu00}. We can 
support this assertion since our numerical model, without the microturbulent 
component, produces no such broadening. Thus, {\it the unrealistic component 
of Z92's model may have been the addition of a microturbulent component, not
the underlying LP flow regime.}

Our main result is that the infall speeds derived from the line profiles 
using simple models are systematically lower than the actual maximum speeds 
arising in the numerical model. This mismatch 
occurs because the outside-in radial velocity profile that self-consistently 
develops in the simulation is very different from the inside-out collapse assumed by 
\citet{Shu77}, which has often been used as a template radial velocity 
profile for the interpretation of line profiles. Instead, in outside-in collapse, 
the highest velocities occur in the core's envelope during the prestellar stage; 
that is, at radii where the density is decreasing, while the densest parts of the
core coincide with the lowest infall speeds further inwards. Therefore,
since the optically thin parts of line profiles are essentially
density-weighted velocity histograms \citep[e.g.,] [] {Pichardo+00}, the
largest velocities are down-weighted by the lower densities, thus
represented by reduced emission within the line, while lower velocities
have a larger weighting and dominate the blue and red peaks, as
well as the absorption dip of the profile.

Quantitatively, the difference between the actual peak velocity for each snapshot in 
Table~\ref{tab:phystimestep} and the corresponding $v_{\rm in}^{Hill5}$ (from figs. 6 
and 7) for the central position ranges from (earliest-latest): 1.5-.3.5 
($\theta_{\rm b}=0.015$~pc), 1.5-3.8 ($\theta_{\rm b}=0.03$~pc) and 1.6-4.5 
($\theta_{\rm b}=0.06$~pc) for HCO$^+$ $J=1-0$; and 1.4-2.3 ($\theta_{\rm b}=0.015$~pc), 
1.5-2.7 ($\theta_{\rm b}=0.03$~pc) and 1.5-3.0 ($\theta_{\rm b}=0.06$~pc) for 
HCO$^+$ $J=3-2$. From this, we can deduce that higher transitions are more accurate in 
estimating the actual core velocity than lower transitions for a given species (see also 
\S5.3) and that the variation between $v_{\rm in}^{Hill5}$ and the actual peak velocity 
in the core increases as the core evolves. Also, larger beams dilute the measured signal 
resulting in lower values of $v_{\rm in}^{Hill5}$.

Moreover, we find that the inferred infall speeds appear to increase as the 
LOS intersects the core closer to its center. This is probably due to the 
fact that, for LOSs farther from the center, the LOS component of the radial 
infall motions is smaller \citep{Anglada+87}. This is consistent with 
interferometric observations of actual cores displaying extended inward motions
\citep[e.g. L1544, L694-2 and NGC 1333][]{taf98,will06,Walsh+06}, which show that the
inward speeds in such cores decrease with increasing projected radius.

We also note that we did not have to consider a static envelope in
order to create the self-absorption. Instead, in our case the
self-absorption is produced by the core's centermost regions (see right panel 
of fig.~\ref{fig:explanopt}), which have the lowest velocities and the highest 
opacities. This is evidenced by the fact that we are only considering half of 
the numerical box for the generation of the line profiles using MOLLIE, so that 
the borders of the region considered have velocities that are typically not less 
than half the maximum infall speeds (see the top panel of fig.\
\ref{fig:physrads}, where the borders of the region considered for the
radiative transfer calculation are marked by vertical dashed lines), so 
that this material cannot be causing the absorption of the near-zero velocity 
emission.

We should also note that, due to the larger central densities towards
the end of the simulation, the self-absorbed spectrum begins to
saturate for spatial positions close to the core center. It becomes
broader and diminishes relatively in brightness compared to 
spectra observed towards neighboring less dense positions. 
This spectrum therefore departs from the typical infall signature that 
is apparent at spatial positions largely separated from the center (e.g. 
compare the spectral panels for snapshot 61 in fig.~\ref{subfig:ts61strip}). 
An attempted Hill5 model fit contains spurious fitted points resulting in a
relatively poorer approximation to the underlying saturated spectrum
and hence a higher $NRMSD$--value. In actual spectral data, low-lying
rotational transitions naturally saturate when the density far exceeds
their $n_{\rm cr}$ for excitation, especially for narrower
$\theta_{\rm b}$, and so too depart from the typical infall
signature. Thus, we infer that low-$n_{\rm crit}$ transitions
suffering from saturation may prove inadequate for correctly capturing
the infall process.

For small radial offsets in fig.~\ref{fig:vin_plot}, corresponding to
$\theta_{\rm b}=0.015$~pc (green datapoints), the derived infall
velocities drop abruptly relative to those further out as a result of
the sensitivity of the Hill5 algorithm to the line profile shape -
that is, its inability to fit saturated lines. Comparing the
resulting fitted data values with the corresponding values for HCO$^+$
$J=3-2$ in fig.~\ref{fig:vin_plot32}, it can be seen that saturated
lines are unreliable when trying to derive infall
velocities. Saturation is lessened by increasing the width of the
synthetic beam by causing a greater fraction of the infalling material 
to be sampled. Analogously, far from the core center, 
the relative shallowness of the
self-absorption dip becomes more prominent as the $T_{\rm b}$/$T_{\rm
r}$--ratio declines, again resulting in an artificial drop in 
$v_{\rm in}^{Hill5}$. The point where this appears is
progressively further away from the center as the core evolves and is
reflected in the number of datapoints included for each snapshot. As
can be seen from the corresponding sets of plots for the two
transitions, the magnitudes of the derived velocities differ on
account of a number of factors: choice of molecular transition
(i.e. $n_{\rm cr}$), proximity to core center and the evolutionary
state of the core \citep[see also][]{keown16}. Also, from
figs.~\ref{fig:vin_plot} and \ref{fig:vin_plot32}, a successively wider 
$\theta_{\rm b}$ results in a slightly reduced $v_{\rm in}^{Hill5}$ for 
small impact parameters relative to the core center. This effect is due to 
larger beams sampling a greater fraction of relatively slower moving 
gas and lessens further from the center as a result to the relative drop
in velocity at all positions for large impact parameters. 

In all modeled spectral positions, the weaker red peak never resembles
a red shoulder to the stronger blue peak, and is nonetheless
significant. The red shoulder feature in line spectra has been associated 
with very large infall velocities \citep[so-called ``fast'' infall,][DM05]{Myers96} 
and is ill-fitted by the Hill5 model. Thus, the large $NRMSD$--values are solely 
due to saturation in the modeled synthetic spectral data. The range of 
differences between the respective infall velocities determined using 
the different values of $\theta_{\rm b}$ is (excluding errors) 
$0.003~\lesssim~\delta v_{\rm in}^{Hill5}~\lesssim~0.05$~$\kms$, 
from non-saturated spectral positions and amongst all analysed snapshots.
Among the different values of $\theta_{\rm b}$, for each transition and 
snapshot investigated, we find that the derived $v_{\rm in}^{Hill5}$
does not vary appreciably for non-saturated ($\sim15$\%), in reasonable 
agreement with DM05.

\subsection{Analytical solutions to the spherical collapse problem and the 
origin of the infall profiles in prestellar cores}
\label{sec:origin}

Our main result (c.f. \S\ref{sec:speed_compar}) is that
standard analysis techniques of the infall profiles in moderately
optically thick lines of collapsing cores tend to systematically
underestimate the speeds occurring in the cores. This result has
profound consequences on our understanding of the dynamical state of
prestellar cores, and thus it is important to understand the origin of
this effect.

\subsubsection{Discussion of spherical collapse models}
\label{sec:collapse_models} 

In order to understand the reason for this sytematic underestimation of
the infall speed by standard analysis methods of line profiles, it is
necessary to discuss the nature of the velocity profile arising in our
simulation, in comparison to the standard ``inside-out'' velocity
profile assumed in most models for infall line profiles.

The velocity profile arising in our simulation is of the ``outside-in''
kind \citep[e.g.,] [Paper I] {GO09, GO11}. By ``outside-in'' we refer
to a flow described by ``Band 0'' in the parameter-space analytical
study of \citet[] [hereafter WS85] {WS85}, which corresponds to strongly
gravitationally unstable initial conditions. This class of collapse flow
includes the standard Larson-Penston solutions \citep{lar69,
pen69}, and is appropriate for our simulation, in which the whole
numerical box is strongly gravitationally unstable. This type of flow is
characterized by an asymptotic solution consisting of a central part
with a roughly uniform density profile, surrounded by an $r^{-2}$
power-law envelope, resembling the radial density scaling of a BE
sphere, although with a higher absolute value of the density than that
required for hydrostatic equilibrium \citep[e.g.,] [Paper I]
{keto15}. For this configuration, the infall speed increases linearly
with radius at the central regions and approaches a supersonic constant
value at the outer envelope. Indeed, the numerical simulation approaches
this regime (see fig.\ 2 of Paper I). The transition between the core
and the envelope occurs at a radius of the order of the Jeans length for
the central density and temperature \citep{KC10}.

This flow structure is in sharp contrast with the standard assumption
that the velocity profile has an inside-out nature \citep[e.g.,] []
{Shu77, Evans99}, in which the infall speed is maximum at the center,
has an $r^{-1/2}$ radial dependence, and extends only to a rarefaction
front located at a radius $\Rshu = \cs t_{\rm PF}$, where $ t_{\rm PF}$
is the time {\it since the formation of the singularity} (the
protostar). Beyond this radius, the inside-out Shu profile assumes that
the gas is still in a hydrostatic state. This radial velocity
profile is applicable only for the protostellar stage (i.e., {\it after}
the formation of the singularity---the protostar), and only for an
initial condition given by a SIS at the time of protostar formation. This 
idealized initial condition assumes that the prestellar evolution of the 
core proceeds quasistatically all the way to the formation of the SIS, 
and only becomes fully dynamic after this time.

\citet{Shu77} argued that the SIS hydrostatic initial condition should be
possible as long as the flow is subsonic, so as to allow the
establishment of detailed pressure balance at all radii. Furthermore, he
argued that the initial and boundary conditions of the Larson-Penston
similarity solutions were highly numerically {\it ad-hoc}; i.e., chosen
to match their numerical results. Furthermore, he argued that the
smoothly (i.e., without a shock), monotonically-decreasing velocity
profile towards the center could only be made consistent with the
outer supersonic inflow through ``an artificial arrangement of
self-gravity and pressure gradient''. 

However, shortly thereafter, \citep{Hunter77} showed that numerical
simulations in general seemed to be best described by the Larson-Penston
similarity solution. The simulation from Paper I, which starts with a
generic Gaussian fluctuation superposed on a uniform medium also
approaches this type of flow. Moreover, \citet{Whitworth+96} and
\citet{VS+05} have later suggested that actually it is the SIS that is
most unlikely. This is because the SIS is an {\it unstable} equilibrium
solution, and moreover, any previous equilibrium solution of the form of
a truncated BE sphere with a central-to-peripheral density contrast
larger than the critical value of $\sim 14$ is unstable as well
\citep{Ebert55, Bonnor56}, so that the SIS is actually the most possibly
unstable case of the family of hydrostatic solutions represented by the
Lane-Emden equation. As such, the SIS as a quasistatic initial condition
for spherical collapse is unrealizable in practice.

It is nevertheless very important to note that the infall velocity 
ideally remains constant at arbitrarily large distances from the core's
center in the outside-in collapse of a uniform-density medium. In
practice, however, this asymptotic solution cannot occur, because the
cloud must end at some point, and the density must decrease again, or at
least be characterized by a density gradient, albeit perhaps weaker than
that in the core. Moreover, the analytic solution \citep{WS85} does not
consider the fact that the fluctuation may be embedded in an already
contracting cloud, so that the gas flows towards the large-scale
collapse center (different from the core's center), as in the
``conveyor-belt'' flows observed in realistic cloud simulations
\citep{GV14} and in the Central Molecular Zone of The Galaxy
\citep{Longmore+14}. On top of this flow, the density fluctuation
(perhaps produced by turbulence) eventually becomes unstable as the
average Jeans mass in the whole cloud decreases due to the global
contraction, and begins to collapse towards its own center. In our
simulation, the decrease of the infall speed towards the edge of the box
is an artifact of the periodic boundary conditions, but this mimics the
effect seen in large-scale simulations that the local collapses within
the cloud extend to finite distances only, because of the finite size of
the turbulent density fluctuations.

In any case, the extent of the collapsing region in the outside-in
collapse case is {\it much} larger than the extent of the standard rarefation 
front of Shu's (1977) inside-out collapse, $\Rshu$ (see above). This is because 
the rarefaction front in Shu's inside-out collapse has not even started to
propagate over the entire duration of our simulation, since it precisely starts 
at the ending time of the simulation (i.e., at the time of formation of the 
singularity). {\it This explains the ``extended inwards motions'' observed in 
actual cores, which extend far beyond the position consistent with Shu's rarefaction
front.}

\subsubsection{The formation of the line profile for outside-in velocity
profiles}
\label{sec:interp_profile} 

\begin{figure*}%
\centering
\label{fig:first}%
\subfigure[Inside-out collapse picture of line asymmetry]{%
\includegraphics[width=0.465\textwidth]{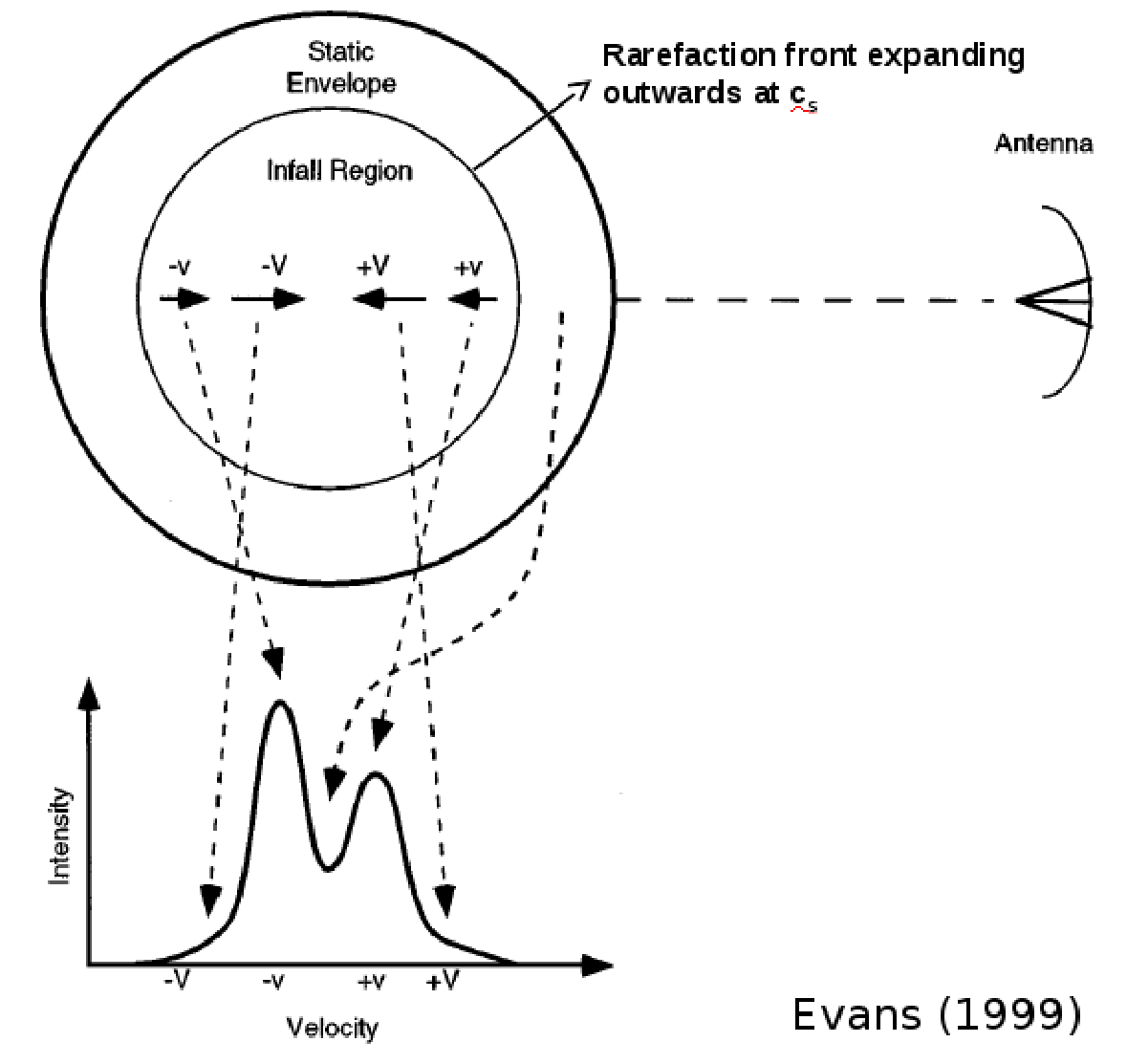}}%
\hspace{0.5cm}
\subfigure[Outside-in collapse picture of line asymmetry]{%
\includegraphics[width=0.5\textwidth,trim=0 -4.0cm 0 0]{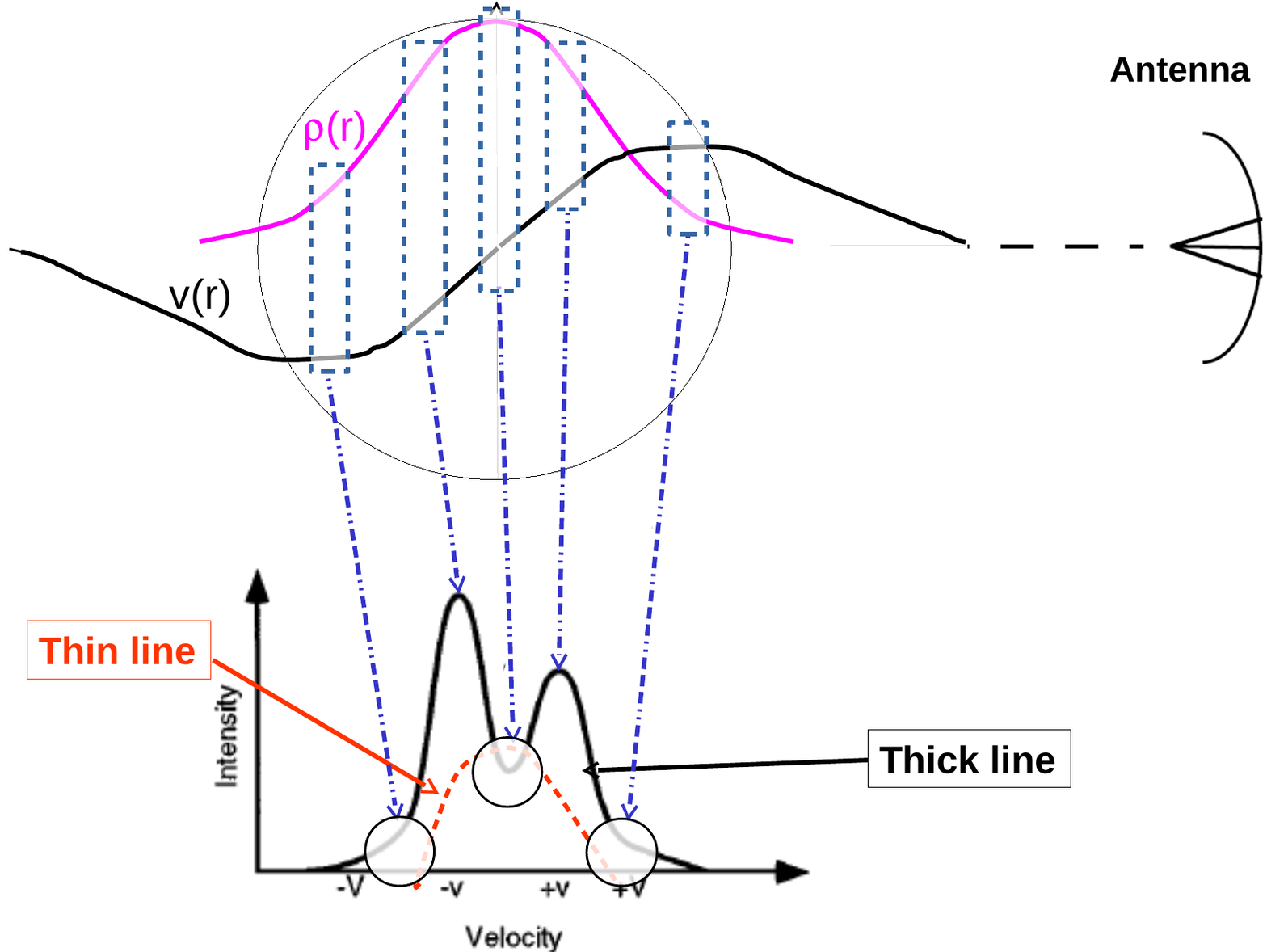}}%
\label{fig:second}%
\caption{Schematic representations of how various parts of the
star-forming mechanisms of (a) inside-out \citep[from][]{Evans99} and 
(b) outside-in collapse (\textit{this work}) contribute to the 
self-absorbed optically thick line profile. In (b), we highlight, using 
rectangular sections of $\tau_{\nu}\sim1$, specific parts for a given snapshot 
of each of its density (\textit{magenta}) and velocity (\textit{black}) profiles 
that give rise to characteristic features of the self-absorbed line. With time, 
the peak velocity increases and is more centralized with the mean velocity of the 
material, responsible for the line self-absorption, shifting to 
redder velocities and therefore displacing the self-absorption minimum to these velocities too. 
Also in (b), a representative optically-thin line is included as emanating from the center 
of the collapse for a high $n_{\rm crit}$ transition. In each panel, the antenna is 
positioned to the right with its LOS passing through the circular core from right to left.}\label{fig:explanopt}
\end{figure*}

In the left panel of fig.~\ref{fig:explanopt}, we show the
``standard picture'' \citep{Evans99} to explain the self-absorbed line 
profile signature on the basis of the inside-out collapse model of 
\citet{Shu77}. In the right panel of this figure, we show the corresponding 
schematic diagram reflecting our interpretation of this phenomenon, on the 
basis of the outside-in collapse characteristic of our numerical simulation. 
In each panel, the observing antenna is located to the right of the schematic
with its LOS passing through the circular core from right to left.

As seen in the left panel of fig.\ \ref{fig:explanopt}, the
structure of the core assumed by \citet{Evans99} is that of Shu's (1977)
inside-out collapse, with a central infalling core and an outer
hydrostatic envelope, mediated by an expanding rarefaction front located
at radius $\Rshu$ (cf. \S\ref{sec:collapse_models}). In this
picture, the central absorption dip in the line profile is caused by the
outer static envelope.  

Instead, as shown by the right panel of fig.\
\ref{fig:explanopt}, for the outside-in profile, there is no
outer hydrostatic envelope, and the outer envelope infalls at speeds 
ranging from the maximum to roughly half that value (cf. dashed portions of 
\ fig.\ \ref{fig:physrads}). In this case, the central absorption dip is produced 
by the high-opacity, low velocity material near the core center, 
while the envelope's velocity appears in the line profile wing, 
downweighted by the lower density there. As mentioned in \S\ref{sec:disc}, 
the fact that the absorption dip is caused by the dense central parts of 
the core is evidenced by the fact that we are only considering the central 
half of the simulation in each direction, so that the gas is still moving 
with supersonic infall velocities at the boundary of this region. Thus, the
only material with near-zero velocity is at the core center.

\subsubsection{Comparison with previous work} \label{sec:compar_prev}

Synthetic observations of various numerical models of
spherically-symmetric collapse have been previously presented by
\citet[] [hereafter K15] {keto15}, to determine which type of
collapse model best matches the observed line profiles of the L1544
starless core. In particular, those authors considered the cases of a
quasi-equilibrium BE sphere (QE-BES), a non-equilibrium BE sphere
(NE-BES), a static sphere, a Larson-Penston (LP) flow, and the
inside-out collapse of an SIS as modeled by \citet{Shu77}. They
concluded that the QE-BES regime produces synthetic line profiles that
best match those observed in the L1544 core.

Interestingly, the evolution of the QE-BES of K15 does have some
similarities with that of our collapsing core embedded in an unstable
uniform medium. The QE-BES model was constructed by K15 as an unstable
BES, and then slightly perturbed, to allow it to evolve hydrodynamically.  
Although K15 refer to the collapse mode of the QE-BES as ``inside-out'', 
it also presents the inner region in which the infall speed {\it increases} 
linearly with radius, so it falls into our description of ``outside-in'' 
collapse. Because of their setup, it also has an outer radius beyond which 
the infall velocity begins to decrease outwards. However, due to the setup 
as a perturbed marginal equilibrium, it develops speeds slightly slower 
than an alternative experiment with an out-of-equilibrium setup, labeled
NE-BES by K15: while the QE-BES shows infall speeds $\sim 0.2\ \kms$
(i.e., $\sim \cs$) at the last timestep shown, the NE-BES develops
speeds $\sim 0.25\ \kms \sim 1.25\ \cs$ at the same time. However, K15
do not indicate how far from the appearance of the singularity are
those times, so it is not possible to know whether supersonic speeds
do appear at later times in the QE-BES case as well.

Similarly to this work, K15 compare optically-thick H$_2$O
($J_{K_{\rm a}K_{\rm c}} = 1_{10} \rightarrow 1_{01}$) and optically
thin C$^{18}$O ($1 \rightarrow 0$) line observations of the L1544 core
against synthetic observations in the same lines of their numerically
simulated cores. They conclude that the QE-BES provides the best match to
the observed profiles. However, K15 concentrate on matching the observed
line profiles, rather than on comparing the actual infall speeds in the
numerical model to the speeds that would be inferred from the line
profiles using standard line-modeling techniques, which is our main
interest here.

K15 devote a full section to justify the feasibility of stable BE
spheres forming within a turbulent molecular cloud environment before
becoming unstable and proceeding to dynamical collapse. They essentially
follow \citet{Field+11} and ``imagine the ISM as a turbulent cascade of
mass and energy from larger to smaller scales'', in which the larger
structures contain supersonic motions and thus fragment, while the
smaller structures (``cores'') are subsonic and thus do not fragment any
further \citep[see also] [] {VS+03}. Then K15 assume that a fraction of
these subsonic cores can be stabilized by the combined effect of thermal
and turbulent energy, and that, as the turbulent energy dissipates, the
cores can go into collapse. They also suggest that cores that are
supported by thermal energy alone may never form stars, and that many of
the cores in the Pipe nebula may be in this category, referring to the
result by \citet{Lada+08} that most of the cores in the Pipe appear to
have masses smaller than their Bonnor-Ebert mass, and therefore must be
gravitationally stable, and confined by an external pressure.

This scenario, however, does not appear feasible in practice. First, as
already stated in \S\ref{sec:backgd}, for a BE sphere to be stable, 
besides having a central-to-peripheral density contrast smaller than the 
critical value, it must be truncated and embedded in a diffuse, warm 
medium, which provides pressure without adding weight \citep{VS+05}. 
Cores deep inside molecular clouds are likely to be embedded 
in the same molecular material as that which they are made of, and therefore 
the tenuous confining medium is not available. In this case, a local 
density enhancement (a ``core'') is also a local pressure enhancement, 
and must therefore re-expand in a sound-crossing time, if it does not 
become locally Jeans-unstable \citep{Galvan+07}. 

Second, in Paper I we showed that our collapsing core, throughout its
evolution, tracks the locus of the Pipe cores in the diagram of
$\Mc/\MBE$ {\it versus} $\Mc$, where $\Mc$ is the core's mass and $\MBE$
is its BE mass, {\it including the region occupied by the apparently
stable cores.} This occurs precisely because our core is just ``the tip
of the iceberg'' of a larger-scale collapse that extends out to the
uniform background. Since this uniform background is generally not
considered part of the core, the core may appear stable, because
not all the mass involved in the collapse is accounted for. The infall
motions that extend into the uniform background constitute an accretion
flow from the cloud onto the core which, however, cannot be described by
a simulation of a core artificially truncated at some radius.

Finally, to our knowledge, no numerical simulation of a turbulent cloud
has ever reported the production of quasi-equilibrium structures. K15
refer to the statement by \citet{Offner+08} that ``the protostellar
cores in the simulations are at the centers of regions of supersonic
infall, which contradicts the observations that show at most transonic
contraction'' \citep[see also] [] {MS13}, suggesting that this may be a
problem of the simulations. However, it should be noted that the
statement by \citet{Offner+08} refers to {\it protostellar} cores, for
which supersonic speeds are more commonly observed, and the discrepancy
they discuss is more quantitative than qualitatve. In fact, those
authors offer the explanation that their simulations lack stellar
feedback that may prevent the development of the very massive stellar
particles, and thus the excessive speeds developing in their
protostellar cores.

\citet{MS13}, on the other hand, do refer specifically to the supersonic
speeds that develop in simulations shortly before the formation of the
protostar (i.e., still during the prestellar stage), and they conclude
that this is indicative of some physical mechanism that is missing from the
simulations, which is required to bring them into concordance with
observations. Our result, that the apparently subsonic nature of the
prestellar collapse may be simply the result of a misinterpretation of
the infall line profiles because of the assumption of an erroneous
infall velocity radial profile, suggests that the problem may lie in the
interpretation of the observations rather than in the simulations.

Instead, our mechanism of global hierarchical gravitational
collapse of MCs \citep[] [see also V\'azquez-Semadeni 2018, in
preparation] {VS+09, BP+18} provides a simple mechanism through which a
density fluctuation (probably of turbulent origin) can at some point
become gravitationally (Jeans) unstable, and begin to
collapse. Succintly, this is just the result of the global reduction of
the average Jeans mass in the cloud as it contracts gravitationally
\citep{Hoyle53}, so that fluctuations of a given mass are stable as long
as their mass is lower than the mean Jeans mass, but, as this mass
decreases over time, they eventually become unstable and begin to
collapse. When this happens, their mass will be just marginally above
the mean Jeans mass in the cloud, similarly to the case of K15's QE-BES.

\subsection{Beamwidth dependence of the profile asymmetry} \label{sec:beam}

We now turn to the origin of the different degrees of asymmetry in 
observational line spectra based on a number of factors including incident 
observational beamwidth, molecular transition and proximity of incident 
beam to core center.
The dependence of the $T_{\rm b}/T_{\rm r}$--ratio on both time and
radial offset from core center is considered in figs.~\ref{fig:tbtr_10}
and \ref{fig:tbtr_32}. In this work, we do not
consider depletion, and allow the emission profiles to develop naturally
for both rotational lines.

As previously discussed, within the quasi-static contraction picture, 
star-forming cores oscillate globally and at the point of gravitational 
instability, the fundamental oscillation mode has zero frequency. 
\citet{stah09,stah10} studied, using perturbation theory, the evolution of a 
3$\Msun$ spherical cloud with such a frozen mode. They found that their cloud 
underwent accelerated, though subsonic, contraction for a period of $\sim$~1~Myr. 
Although their resulting line spectra were mildly asymmetric, they concluded that 
the accelerating character of their model naturally explained why low density 
starless cores exhibit line spectra with smaller $T_{\rm b}/T_{\rm
r}$--ratios. However, unlike in this work, their model was unable to
account for the extended spatial occurrence of asymmetric line profiles
\citep{stah10}.

Instead, in our synthetic observations of the simulated
collapsing core, this ratio is consistently higher for HCO$^+$ $J=3-2$ than 
for $J=1-0$ if we compare corresponding panels in figs.~\ref{fig:tbtr_10} and
\ref{fig:tbtr_32} for small radial offsets from the core center. In
fact, from the panels, where $\theta_{\rm b}=0.06$~pc, the difference
grows as the core evolves. The $J=3-2$ emission is more centrally
confined than the $J=1-0$ emission on account of its larger $n_{\rm
cr}$ (see Table~\ref{tab:moltrans}). Thus, for the more centrally dense 
core snapshots considered here, the increasing infall velocity (as the
core evolves) is more apparent for the higher transition because the
mean density sampled by the respective values of $\theta_{\rm b}$, for
each snapshot, is maintained within its $n_{\rm cr}$ while it exceeds
the $n_{\rm cr}$ of HCO$^+$ $J=1-0$, evidenced by saturation at small
$\theta_{\rm b}$ for the latest snapshots ($T_{\rm b}/T_{\rm r}<1.0$
at $\theta_{\rm b}=0.015$~pc for snapshot 65 in
fig.~\ref{fig:tbtr_10}). Since HCO$^+$ $J=3-2$ does not succumb as
easily to saturation as the $J=1-0$ line, it produces stronger a
$T_{\rm b}/T_{\rm r}$ ratio resulting in a higher derived $v_{\rm
in}$.

\begin{figure}
\includegraphics[width=\linewidth]{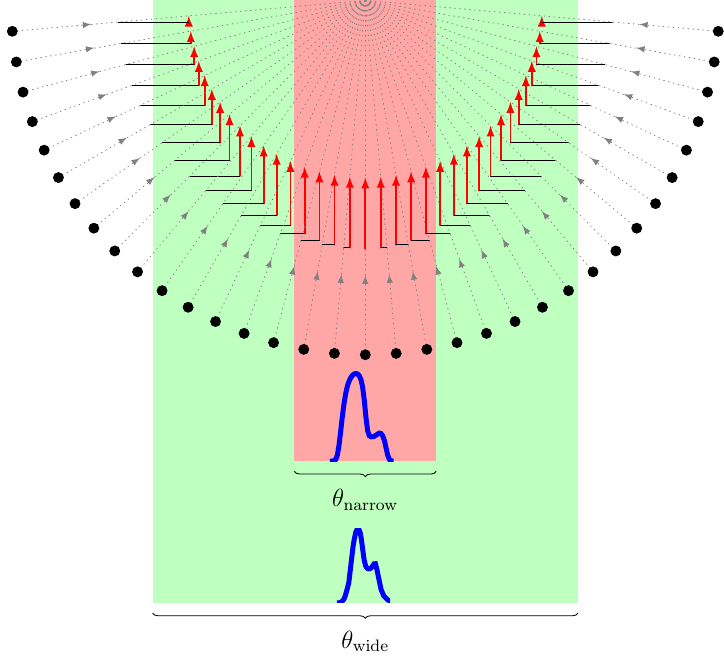}
\caption{Simple diagram explaining the detection of larger $T_{\rm b}/T_{\rm r}$--ratios for collapsing core gas (black dots of equal velocity). 
$\theta_{\rm wide}$ captures a larger fraction, by volume (in green), of the emission from infalling gas along the LOS through the core 
center, while $\theta_{\rm narrow}$ captures a smaller, localized volume (in pink). From the diagram, 
$\theta_{\rm narrow}$ traverses infalling gas traveling mainly parallel to the LOS (velocity components in red). $\theta_{\rm wide}$ traverses 
this same gas as well as gas oriented at larger angles (smaller red components) relative to the LOS. The resulting beam 
convolution for this detected gas of similar velocity will result in larger $T_{\rm b}/T_{\rm r}$--ratios for $\theta_{\rm narrow}$. This holds 
similarly for the detection of large $T_{\rm b}/T_{\rm r}$--ratios far from
the core center using a wider beam.}
\label{fig:planar}
\end{figure}

For each transition, at small offsets from the center, $T_{\rm
b}/T_{\rm r}$ is seen to decrease as $\theta_{\rm b}$ is increased 
(for {\it non-saturated} positions). The common explanation for this, 
illustrated in fig.\ \ref{fig:planar}, is that a wider beam passing through 
the core's center picks up emission from a larger region, but in which the 
additional material is moving more obliquely with respect to the LOS 
\citep{Anglada+87, Anglada+91}, and therefore the projection of its velocity 
onto the LOS is smaller. Thus, in a wider beam passing through the center, 
most of the gas contributes lower velocities, and therefore the 
$T_{\rm b}/T_{\rm r}$ ratio is also smaller while, conversely, a more compact 
sampling of the collapsing gas produces a more prominent line asymmetry, 
especially closer to the core center, again resulting in a higher derived 
$v_{\rm in}$ (see also figs.~\ref{fig:vin_plot} and \ref{fig:vin_plot32}). 
At large radial offsets from the core center using both lines, the value of
$T_{\rm b}/T_{\rm r}$ falls between 1.1 and 1.35 depending on the stage
of collapse and the value of $\theta_{\rm b}$. These positions coincide
with relatively quiescent core material with their $T_{\rm b}/T_{\rm
r}$--values giving the impression of the presence of a static ($T_{\rm
b}/T_{\rm r}\approx$~1.0) external envelope relative to the more dynamic
infalling core center.

The $T_{\rm b}$/$T_{\rm r}$ ratio is therefore very sensitive to the
excitation conditions in the core. Better estimates of core infall
motions are possible by observing the higher rotational transitions of
commonly-observed molecular species using as finite a beam as
possible. 

Another very noteworthy point to make is that our synthetic spectra for
the HCO$^+$ $J=3-2$ transition reproduce the so-called extreme $T_{\rm
b}$/$T_{\rm r}$ ratios ($\gtrsim$~3, see last two panels of
fig.~\ref{fig:tbtr_32} for snapshot 65 with $\theta_{\rm b}=0.03$ and
0.06~pc respectively), seen in the evolved collapsing collapsing sources
B133 \citep{greg00a} and NGC 7538 \citep{sun09} using similarly
optically thick transitions. Since each of these cores harbors a central
source (both are Class 0 cores), while snapshot 65 of the simulation 
represents the end of the simulated prestellar core collapse, at which 
point a central source is about to form, their respective velocity profiles 
should resemble one another (e.g. see fig.~\ref{subfig:radialvel}). We 
therefore suggest that these particular cores are undergoing a similar 
physical process to that exhibited by the final stages of our simulation.

\subsection{Radial variation of $\delta v$}

For the asymmetry parameter ($\delta v$), in both transitions, the
general consensus is that narrower beams result in smaller
$\left|\delta v\right|$ at the core center but with larger values far
from the center for a given snapshot. For the later stages of
collapse, prior to star formation, this parameter tends to increase
with increasing radial offset from the core center due to the decrease 
in $\Delta v_{\rm thin}$ and the inverse relationship between 
$\left|\delta v\right|$ and $\Delta v_{\rm thin}$. In view of the results 
presented here, the strict $\left|\delta v\right|>0.25$
requirement for the assignment of true collapsing motions (see
Mardones et al. 1997) would need to be revised for mapped observations
of evolved collapsing cores, especially for typical values of
$\theta_{\rm b}$, given how $\left|\delta v\right|$ behaves very close
to evolved collapse centers where $\Delta v_{\rm thin}$ is largest.

At the core center for snapshot 49, the optically thick peak for
HCO$^+$ $J=3-2$ is shifted blueward when compared to the same region
in snapshot 52 (top two panels of fig.~\ref{fig:asymplot2}) resulting
in larger values for $\left|\delta v\right|$. Since HCO$^+$ $J=1-0$
does not exhibit this feature at any value of $\theta_{\rm b}$, this
may indicate that the higher transition is more sensitive to the
velocity field at comparatively lower densities. Differing velocity
field gradients in the detected emission lead to variations in the
position of the emission peak. Narrower beamwidths are more sensitive
to such variations and so a $\delta v$--analysis of mapped spectral
regions can be used to identify gradients in the velocity field.

Given the form of eq.~\eqref{eq:asympar}, a numerical analysis of this
parameter is limited by the velocity resolution of the radiative
transfer setup. Nevertheless, we have been able 
to make some useful deductions from such an analysis of the synthetic 
spectra.

\subsection{Radial variation in the optically thick line}

It is also worth noting that our simulated core exhibits a behavior in
the optically thick \hcop\ $J=3-2$ line that is qualitatively very
similar to that observed by P10 for \nht (1,1) lines from the B5-1
cloud core as the impact parameter of their LOSs was increased (see
their fig.\ 3 and compare to our fig.\ \ref{fig:specpanels}). These
authors noted that the \nht\ line towards this core, with its 
blue-skewed, double-peaked asymmetry, became weaker and its central 
dip tended to disappear as the impact parameter was increased. This 
``line'' actually consists of 8 magnetic hyperfine 
components due to the combined coupling of the magnetic hyperfine 
interaction of the three identical $^1$H-nuclei and the quadrupolar 
moment of the $^{14}$N-nucleus in \nht\ \citep[e.g., see] [] {Ku67}.
  
P10 interpreted the combination of the two peaks as the blending of
these underlying hyperfine components. Using those components, they
fitted the central \nht\ line for several individual pointings observed
towards the core. From their fit, they concluded that the linewidth of 
each of the hyperfine components in the fit decreased as the impact 
parameter decreased---an indication of a ``transition to coherence'' 
(i.e., a reduction of the turbulent velocity dispersion) as the 
impact parameter decreased. Further, they suggested that the gradual 
merging of the two peaks as the impact parameter is increased is a 
manifestation of an increasing turbulent component that becomes wider than 
either of the two peaks, eventually incorporating both of them at sufficiently 
large values of the impact parameter. However, it is noteworthy that the 
overall linewidth of the line does not vary -- the interpretation of an
increasing turbulent component at higher impact parameters was based
exlusively on the gradual disappearance of the central dip and the
peaks, leaving behind a single line that has approximately the same
width as the double line seen at smaller impact parameters.

On the other hand, our \hcop\ line does not contain a discernible
hyperfine structure unlike \nht, and our simulation does not include
turbulence, yet the observed behavior of the line is qualitatilvely very
similar to that reported by P10. In our case, the interpretation must be
that, as the impact parameter of the LOS is increased, the LOS ceases to
pick up the high-density, low-infall velocity at the center, which makes
up the central absorption dip, and only picks up the higher-velocity 
infalling gas, although at a higher angle with respect to the LOS, so the 
net velocity traced by the linewidth remains nearly constant. Also, the 
absolute intensity of the line decreases because of the drop in density at 
larger impact parameters. In a future contribution we will examine in detail 
the \nht\ line with its hyperfine structure, but it is tempting to speculate 
that the observed behavior of the \nht\ line reported by P10 may be due to
the process we describe, rather than to a transition to coherence.

\section{Summary and Concluding Remarks} \label{sec:concls}

We have produced synthetic spectra for the \hcop\ $J=1-0$
and $J=3-2$, as well as the \nhp\ $J=1-0$
transitions for a number of snapshots of an idealized numerical
simulation of the collapse of a gaussian density fluctuation with
spherical symmetry immersed in a strongly gravitationally unstable
background medium, within the context of the hierarchical gravitational
collapse of molecular clouds \citep{VS+09}. The simulation was presented
in \citet[] [Paper I] {rn15}, and started from a minor, Jeans-sized
fluctuation of amplitude 50\% above the density of the uniform
background, so the density and velocity fields of the simulation
developed self-consistently. We examined 7 snapshots covering the time
elapsed since the density fluctuation had a density contrast of $\sim
3$ with respect to the background to when this contrast was $\sim
2000$. 

The main motivation for the present study was the discrepancy between
the supersonic infall velocities that develop in this kind of model
\citep [e.g.,] [] {lar69, pen69} and the subsonic infall speeds that are
often inferred from the analysis of self-absorbed, optically-thick lines
that exhibit a blue excess \citep[e.g.,] [] {zhou92, zhou93} and from
the linewidths measured at distances $\lesssim 0.1$ pc in dense
molecular cores \citep[e.g.,] [] {Goodman+98}. This discrepancy has led
to the general belief that dynamic collapse solutions of the
hydrodynamic equations do not correctly represent the state of such
cores, and that instead the cores need to be supported against their
self-gravity by some agent like magnetic fields \citep[e.g.,] [] {shu87}
or turbulence \citep[e.g.,] [] {MK04, BP+07}. 

However, one possible resolution of the discrepancy may lie in the very
different nature of the generally assumed infall velocity profile for
the interpretation of the lines and the one that develops in collapse
simulations. Often, infall profiles are modeled with the assumption of a
\citet{Shu77} ``inside-out'' profile, which has the highest velocities at
the innermost regions of the cores and has zero velocity (i.e., remains 
hydrostatic) at large radial distances. This profile arises from the 
assumption of a singular isothermal sphere as an initial condition of the 
collapse. Instead, relaxation of this \citep[unrealistic; e.g.,] []
{Whitworth+96} assumption generally produces ``outside-in'' velocity
profiles, for which the maximum speeds occur at a finite distance from
the core's center (specifically, in the core's envelope) during the
prestellar stage \citep[e.g., ] [Paper I] {WS85, Gomez+07, GO11}. In
Paper I we speculated that this property of the infall velocity profile
might cause a systematic underestimation of the infall speeds, because,
noting that molecular line profiles are essentially density-weighted 
histograms of the LOS component of the velocity, such a combination 
of density and velocity profiles might cause the highest velocities to be 
down-weighted by the fact that they occur in the regions of the core where 
the density is already decreasing.

Our synthetic observations, and their analysis using the Hill5
method from \citet{dev05}, confirm the speculation from Paper I,
showing that the speeds inferred from the profiles underestimate the
actual peak infall speed by factors of 2--4, thus creating the
appearance of a subsonic collapse in spite of the presence of supersonic
velocities in the core. This is because, contrary to the standard assumption of 
an inside-out infall velocity profile, our collapsing core develops an outside-in 
profile, in which the largest velocities occur where the density is already 
decreasing, thus downweighting the contribution of the fastest-moving material 
to the line profile. Instead, the higher weighted speeds are the inferior ones 
occuring near the core center, where the density is highest. It is important to 
remark, however, that the usage of the Hill5 method was not essential to our 
results: the fact that the profiles would in general be interpreted as implying
subsonic speeds is clear, simply by the often-used zeroth-order approximation
of measuring the velocity difference between the blue peak and the absorption 
dip, which in all of the spectra shown in fig.~\ref{fig:specpanels} is less than 
the sound speed. 

We also found from the sets of $T_{\rm b}$/$T_{\rm r}$ radial profiles for the \hcop\ $J=3-2$ 
line that this simple collapse simulation reproduces the large $T_{\rm b}$/$T_{\rm r}$--ratios 
observed towards a number of evolved low-mass starless and protostellar cores, contrary 
to the common belief that these large ratios cannot be obtained with simple models \citep{greg00}. 
By accommodating a more extended collapse incorporating both the inner core material 
and the outer envelope, comparatively larger infall velocities are developed in the later stages 
of this simple simulation than in previous setups. The simultaneity of a more extended
infall pattern along the LOS together with larger magnitude infall velocities therefore enables
the development of larger $T_{\rm b}$/$T_{\rm r}$--ratios.

The magnitude of the asymmetry parameter, $\left|\delta v\right|$, increases with 
radial offset at all times (see fig.~\ref{fig:asymplot1} and \ref{fig:asymplot2}). 
This trend is most probably associated with the emerging velocity profile at these times 
in the simulation and, in particular, how this profile effects the value of 
$\Delta v_{\rm thin}$ in eq.~\ref{eq:asympar}. Additionally from fig.~\ref{fig:asymplot1}-\ref{fig:asymplot2}, 
it can be seen that, for smaller $\theta_{\rm b}$, $\left|\delta v\right|$ 
is larger at small radial offsets from the center and decreases far from the core center for 
both HCO$^+$ transitions. This is due to a narrowing of $\Delta v_{\rm thin}$ with distance 
from the core center, especially for later times. Saturation effects result in $|\delta v|<0.25$ for $J=3-2$ at 
later times, especially for spatial positions close to the core center. As we discussed in 
relation to the analytically derived $v_{\rm in}$ values for both $J=1-0$ and $3-2$ at later 
times, saturation effects appear more severe for the higher transition and are persistent 
even at larger values of $\theta_{\rm b}$. From this analysis, we can say that $\delta v$ 
is strongly dependent on the underlying velocity profile along a given LOS due to the 
sensitivity of $\Delta v_{\rm thin}$.

Finally, we also noted that the shape of the synthetic \hcop\ $J=3-2$
line profiles varies with increasing offset in a qualitatively
similar way to that reported by \citet[] [P10] {pine10}, with the
absorption dip becoming less pronounced, while the overall width of the
line is unchanged. While P10 interpreted this as the consequence of the
broadening of a hypothetical turbulent component towards large impact
parameters that causes a blending of the blue and read peaks, in our
case the lack of turbulence prevents this interpretation from being
applicable. Instead, we interpreted the effect as the consequence of the
fact that, as the impact parameter of the LOS increases, the LOS ceases
to pick up the high-density, low-infall velocity at the center, which
makes up the blue and red peaks and creates the self-absorption dip at
the center, and only picks up the lower-density, higher-velocity, more
distant infalling gas. In a future contribution we hope to investigate
whether this scenario is applicable to \nht\ synthetic observations of
our core.

We conclude that the apparently subsonic velocities
generally inferred for low-mass cores, and the standard inter-
pretation that they imply that the cores are not collapsing,
and supported by either turbulence or magnetic fields, may
be an artifact of the outside-in nature of the velocity profile
that naturally arises in collapsing prestellar cores that start
from realistic initial conditions, because this profile implies
that the largest velocities are given less weight in the process
of line formation, since they occur at large distances from
the core center, where the density is lower than at the center.
In addition, we have speculated that the offset dependence
of the infall line profiles, with a disappearing central dip at
larger offsets, may be also a consequence of the outside-in
radial velocity profile rather than an indication of a ``tran-
sition to coherence''.

We have carried out an in-depth synthetic spectral analysis for the simple 
model at our disposal. Factors such as the influence of the background density 
on the depth of the self-absorption dip and the use of different abundance 
profiles could be considered for future work but do not add to the context 
of this paper.

\section*{Acknowledgments}

We gladly acknowledge useful discussions with Susana Lizano that led us
to investigate the relation between the observed and the intrinsic
infall speeds. We also thank Roberto Galv\'an-Madrid for some helpful
suggestions regarding the creation of the synthetic spectra using
different beamwidths. Neal J.\ Evans and Phil Myers greatly aided in our
understanding of which pairs of transitions to use in the determination
of the asymmetry parameter and regarding the radiative transfer
modelling. The numerical simulation was performed in the cluster
acquired with CONACYT grant 102488 to E.V.-S. Also, R.M.L.\ was
partially supported with funds from this grant.

\appendix

\label{lastpage}

\end{document}